\RequirePackage[hyphens]{url}  
\documentclass[sigconf]{acmart}

\pdftrailerid{} 
\setcopyright{none}
\renewcommand\footnotetextcopyrightpermission[1]{}
\acmConference[~]{}{}{}

\usepackage{url} 

\usepackage{hyperref}         
\hypersetup{ 
colorlinks=true, 
breaklinks=true,
linkcolor=black,
citecolor=black,
urlcolor=black,
colorlinks
}
\usepackage{amsmath}          
\usepackage{amssymb}          
\usepackage{amsfonts}         
\usepackage{anyfontsize}      

\usepackage{soul}
\setuldepth{Berlin}

\usepackage{subfig}			  
\usepackage{algorithm}
\usepackage{algpseudocode}

\usepackage{array,multirow,multicol,ragged2e}     
\usepackage{tabularx}                             
\usepackage{threeparttable}                       
\usepackage{stfloats}

\usepackage{graphicx}
\graphicspath{{img/}}

\usepackage{xspace}           

\usepackage{xparse}           

\usepackage{xcolor,colortbl}  
\usepackage{hhline}           
\usepackage{listings}         
\usepackage{color}

\usepackage{enumitem}         

\usepackage{xifthen}
\usepackage{tikz}
\usetikzlibrary{arrows, calc, decorations.markings, positioning}

\definecolor{editorGreen}{rgb}{0, 0.5, 0} 
\definecolor{brown}{rgb}{0.3,0.15,0}
\colorlet{punct}{red!60!black}
\definecolor{background}{HTML}{EEEEEE}
\definecolor{delim}{RGB}{20,105,176}
\colorlet{numb}{magenta!60!black}

\lstdefinelanguage{HTML5}{
	language=html,
	sensitive=true,   
	alsoletter={<>=-},    
	morecomment=[s]{<!-}{-->},
	tag=[s],
	otherkeywords={
		>,
		<!DOCTYPE,
		</html, <html, <head, <title, </title, <style, </style, <link, </head, <meta, />,
		</body, <body,
		</div, <div, </div>, 
		</p, <p, </p>,
		</script, <script,
		<canvas, /canvas>, <svg, <rect, <animateTransform, </rect>, </svg>, <video, <source, <iframe, </iframe>, </video>, <image, </image>, <header, </header, <article, </article
	},
	ndkeywords={
		=,
		charset=, src=, id=, width=, height=, style=, type=, rel=, href=, data-type=
		fill=, attributeName=, begin=, dur=, from=, to=, poster=, controls=, x=, y=, repeatCount=, xlink:href=,
		margin:, padding:, background-image:, border:, top:, left:, position:, width:, height:, margin-top:, margin-bottom:, font-size:, line-height:,
		transform:, -moz-transform:, -webkit-transform:,
		animation:, -webkit-animation:,
		transition:,  transition-duration:, transition-property:, transition-timing-function:,
	}
}
\lstdefinestyle{mylisting} {%
	basicstyle={\footnotesize\ttfamily},   
	frame=none,
	xleftmargin={0.1cm},
	xrightmargin={0.1cm},
	numbers=left,
	stepnumber=1,
	firstnumber=1,
	numberfirstline=false, 
	numbers=none,
	identifierstyle=\color{black},
	keywordstyle=\color{blue}\bfseries,
	ndkeywordstyle=\color{editorGreen}\bfseries,
	stringstyle=\color{brown}\ttfamily,
	commentstyle=\color{brown}\ttfamily,
	alsodigit={.:;},  
	tabsize=2,
	showtabs=false,
	showspaces=false,
	showstringspaces=false,
	extendedchars=true,
	breaklines=true,
	literate=%
	{Ö}{{\"O}}1
	{Ä}{{\"A}}1
	{Ü}{{\"U}}1
	{ß}{{\ss}}1
	{ü}{{\"u}}1
	{ä}{{\"a}}1
	{ö}{{\"o}}1
}

\lstdefinelanguage{json}{
	literate=
	*{0}{{{\color{numb}0}}}{1}
	{1}{{{\color{numb}1}}}{1}
	{2}{{{\color{numb}2}}}{1}
	{3}{{{\color{numb}3}}}{1}
	{4}{{{\color{numb}4}}}{1}
	{5}{{{\color{numb}5}}}{1}
	{6}{{{\color{numb}6}}}{1}
	{7}{{{\color{numb}7}}}{1}
	{8}{{{\color{numb}8}}}{1}
	{9}{{{\color{numb}9}}}{1}
	{:}{{{\color{punct}{:}}}}{1}
	{,}{{{\color{punct}{,}}}}{1}
	{\{}{{{\color{delim}{\{}}}}{1}
	{\}}{{{\color{delim}{\}}}}}{1}
	{[}{{{\color{delim}{[}}}}{1}
	{]}{{{\color{delim}{]}}}}{1},
}

\newcommand{\mcrot}[4]{\multicolumn{#1}{#2}{\rlap{\rotatebox{#3}{#4}~}}} 

\newcommand{\subhead}[1]{\vspace{2pt}\noindent{\textbf{#1.}}}

\newcommand{\yes}{{\small\checkmark}}

\newcolumntype{L}[1]{>{\raggedright\let\newline\\\arraybackslash\hspace{0pt}}m{#1}}
\newcolumntype{C}[1]{>{\centering\let\newline\\\arraybackslash\hspace{0pt}}m{#1}}
\newcolumntype{R}[1]{>{\raggedleft\let\newline\\\arraybackslash\hspace{0pt}}m{#1}}

\begin{document}
	
\author{Xavier de Carn\'e de Carnavalet}
\email{x\_decarn@ciise.concordia.ca}
\orcid{0000-0003-2664-3963}
\author{Mohammad Mannan}
\email{mmannan@ciise.concordia.ca}
\orcid{0000-0002-9630-5858}
\affiliation{
	\institution{Concordia University}
	\city{Montreal}
	\state{QC}
	\country{Canada}
}

\renewcommand{\shortauthors}{Carnavalet and Mannan}
\title[Privacy and Security Risks of ``Not-a-Virus'' Bundled Adware]{Privacy and Security Risks of ``Not-a-Virus'' Bundled Adware: \\The Wajam Case}

\begin{abstract}
Comprehensive case studies on malicious code mostly focus on botnets and worms (recently revived with IoT devices), prominent pieces of malware or Advanced Persistent Threats, exploit kits, and ransomware. However, adware seldom receives such attention. Previous studies on ``unwanted'' Windows applications, including adware, favored breadth of analysis, uncovering ties between different actors and distribution methods.
In this paper, we demonstrate the capabilities, privacy and security risks, and prevalence of a particularly successful and active adware business: \emph{Wajam}, by tracking its evolution over nearly six years.
We first study its multi-layer antivirus evasion capabilities, a combination of known and newly adapted techniques, that ensure low detection rates of its daily variants, along with prominent features, e.g., traffic interception and browser process injection.
Then, we look at the privacy and security implications for infected users, including plaintext leaks of browser histories and keyword searches on highly popular websites, along with arbitrary content injection on HTTPS webpages and remote code execution vulnerabilities.
Finally, we study Wajam's prevalence through the popularity of its domains.
Once considered as seriously as spyware, adware is now merely called ``not-a-virus'', ``optional'' or ``unwanted'' although its negative impact is growing. We emphasize that the adware problem has been overlooked for too long, which can reach (or even surplus) the complexity and impact of regular malware, and pose both privacy and security risks to users, more so than many well-known and thoroughly-analyzed malware families. 

\end{abstract}

\maketitle

\section{Introduction}
The business of generating revenue through ads can be very intrusive for end users.
Popular application download websites are known to bundle adware with their custom installers~\cite{install-top10,giri2016alerting}. Users can also be misled to install Potentially Unwanted Programs/Applications (PUP/PUA) that provide limited or deceptive services (e.g., toolbars, cleanup utilities) along with invasive ads~\cite{adware-blackhat05,wajam-usenix-google}.
The prevalence of adware is also increasing. Recent studies~\cite{wajam-usenix-imdea,wajam-usenix-google} show that Google Safe Browsing triggers 60 million warnings per week for bundled installers, twice the rate of malware-related warnings. 

However, adware applications are generally not considered as much of a threat as malware---apparent from some antivirus labels, e.g., ``not-a-virus'', ``Unwanted-Program'', ``PUP.Optional'', which may not even trigger an alert~\cite{eset-pup,kaspersky-riskware}.
After all, displaying ads is not considered a \emph{malicious} activity, and users even provide some form of ``consent'' to install these unwanted bundled applications~\cite{wajam-usenix-google}.
However, prior to 2006, adware was also labeled as ``spyware''~\cite{symantec2005}, due to its privacy-invasive nature. Since then, several lawsuits succeeded in downgrading the terms used by AV companies to adware, then to PUP/PUA~\cite{mcfedries2005technically,adware-blackhat05}.
Consequently, adware has received less scrutiny from the malware research community in the past decade or so. Indeed, studies on PUPs tend to focus mostly on the revenues, distribution and relationships between actors~\cite{ad-injection,wajam-usenix-imdea,wajam-usenix-google}, and the abuse of code signing certificates by PUPs to reduce suspicion~\cite{certified-pup}. Recent industry reports are now only focused on more trendy threats, e.g., ransomware, supply chain attacks~\cite{symantec2018}. \looseness=-1

Malware analysis has a long history in the academia---starting from the Morris Worm report from 1989~\cite{morris-worm}.
Past malware case studies focused on regular botnets~\cite{botnet-takeover}, IoT botnets~\cite{understand-mirai}, prominent malware~\cite{conficker-study,zeus-concordia}, web exploit kits~\cite{state-exploit-kits,pexy}, Advanced Persistent Threats~\cite{stuxnet,apt1-mandian}, and ransomware~\cite{cutting-gordian-knot}. Results of these analyses sometimes lead to the identification, and even prosecution of several malware authors~\cite{zeus-author-arrest,mirai-author-arrest}, and in some reduction of exploit kits (at least temporarily, see, e.g.,~\cite{exploit-kits-gone}). However, adware campaigns remain unscathed.
Previous cases of ad-related products received media attention as they severely downgrade HTTPS security~\cite{superfish,privdog}, but they generally do not adopt techniques from malware (e.g., obfuscation and evasion).
Therefore, security companies may prioritize their effort on malware, while academic researchers may consider adware as a non-problem, or simply a technically uninteresting one, enabling adware to survive and thrive for long. 
Important questions remain unexplored about adware, including: 1) Are they all simply displaying untargeted advertisements? 2) Do they pose any serious security and privacy threats? 
3) Are all strains limited in complexity and reliably detected by AVs?

On mobile platforms, applications are limited in their ability to display ads and steal information. For instance, an app cannot display ads within another app, or systematically intercept network traffic without adequate permissions and direct user consent. Apps found misbehaving are evicted from app markets, limiting their impact. Unfortunately, there are no such systematic equivalent on desktop platforms (except Windows 10 S mode), and users must bear the consequences of agreeing to fine print terms of services, which may include the installation of numerous bundled unwanted commercial pay-per-install applications~\cite{wajam-usenix-google}.

We explore the case of \emph{Wajam}, a seven-year old \emph{advertisement-supported} \emph{social search engine} that progressively turned into sophisticated deceptive adware and spyware, originally developed by a Canadian company and later sold to China.
We initially observed TLS certificates from some user machines with seemingly random issuer names, e.g., \texttt{b02669b9042c6a8f}. Some of those indicated an email address that led us to Wajam, and we collected 52 samples dated from 2013 to 2018. Historical samples are challenging to obtain, since Wajam is often dynamically downloaded by other software installers, and relies either on generic or randomized filenames and root certificates, limiting the number of searchable fingerprints.

Wajam probably would not subsist for seven years without affecting many users, and in turn generating enough revenue. To this end, we tracked 332 domain names used by Wajam, 
as found e.g., in code signing certificates, and hardcoded URLs in samples,
and followed the evolution of these domains in top domain lists.
In the past two years, we found ranks as high as the top 29,427 in Umbrella's list of top queried domains~\cite{umbrella-1m}. 
Combined together using the Dowdall rule (cf.~\cite{LePochat2019}), these domains could rank up to the top 5,246.
Wajam's domains are queried when ads are injected into webpages and while pulling updates, suggesting that a substantial number of users remain continuously infected.
Indeed, during an investigation by the Office of the Privacy Commissioner (OPC) of Canada in 2016~\cite{opc-wajam}, the company behind Wajam reported to OPC that it had made ``hundreds of millions of installations'' and collected ``approximately 400 terabytes'' of personal information. 

We study the technical evolution of content injection, and identify four major generations, including browser add-on, proxy settings changer, browser process injector, and system-wide traffic interceptor.
Browser process injection involves hooking into a browser to modify the traffic after it is decrypted and before it is rendered, enabling man-in-the-browser (MITB) attacks.
Such attacks are new in the adware realm---known to be last used by the Zeus malware for stealing banking information~\cite{zeus,zeus-spyeye}.

Across generations, Wajam increasingly makes use of
several anti-analysis and evasion techniques including: a) daily release of metamorphic variants, b) steganography, c) string and library call obfuscation, d) encrypted strings and files, e) deep and diversified junk code, f) polymorphic resources, g) valid digital signatures, h) randomized filenames and root certificate Common Names, i) and encrypted updates.
Wajam also implements anti-detection features ranging from disabling Windows Malicious Software Removal Tool (MRT), self-excluding its installation paths from Windows Defender, and sometimes leveraging rootkit capabilities to hide its installation folder from users.
We detail 23 such techniques, which are still effective as of Apr.\ 2019 to prevent most AVs to even flag fresh daily samples. For example, the sample from Apr.\ 29 is flagged only by 4 AVs out of 71, three of them label it with ``heuristic'', ``suspicious'' and ``Trojan.Generic,'' suggesting that they merely detect some oddities.

We also found security flaws that have exposed (possibly) millions of users for the last four years and counting to potential arbitrary content injection, man-in-the-middle (MITM) attacks, and remote code execution (RCE). MITM attacks could make long-lasting effects by changing Wajam's update URL to an attack server.
As the third generation of Wajam leverages browser process injection, content can be injected in the webpage \emph{without} its HTTPS certificate being changed, preventing even a mindful user from detecting the tampering.
In addition, Wajam systematically downgrades the security of a number of high-profile websites by removing their Content Security Policy, e.g., \emph{facebook.com}, and other security-related HTTP headers from the server's response.
Further, Wajam sends---\emph{in plaintext}---the browsing histories from four major browsers (if installed), and the list of installed programs, to Wajam's operators. Finally, search keywords input on 100 groups of domains spanning millions of websites are also leaked. Hence, Wajam remains as a major privacy and security threat to millions of users. 

While the existence of traffic-injecting malware is known~\cite{zeus,zeus-spyeye}, and TLS flaws are reminiscent of Superfish and Privdog~\cite{superfish,privdog}, Wajam is unique in its sophistication, and has a broader impact. Its anti-analysis techniques became more advanced and innovative over time---posing as a significant barrier to study it. 
We also discovered a separate piece of adware, \emph{OtherSearch}, which reuses the same model and similar techniques as Wajam. This indicates the existence of a common third-party obfuscation framework provider, which perhaps serve other malware/adware businesses. We focus on Wajam only due to the abundance of samples we could collect.
Considering Wajam's complexity and automation of evasion techniques, we argue that adware mandates more serious analysis effort. \looseness=-1

\subhead{Contributions}
\begin{enumerate}[itemsep=2pt,topsep=0pt,leftmargin=.55cm,itemindent=0cm]
\item We collect and reverse-engineer 52 unique samples of Wajam spanning across six years and identify four content injection techniques, one of which was previously used in a well-known banking trojan.
This analysis is a significant reverse-engineering effort to characterize the technical and design evolution of a successful ad injector. We investigate the chronological \emph{evolution} for such an application over the years, shedding light on the practices, history and techniques used by such software. Our analysis may help advance reverse engineering of other malware as well.

\item We uncover the serious level of complexity used in Wajam across generations. These 52 samples used various combinations of 23 effective anti-analysis and evasion techniques, and even rootkit-like features, which are even rarely found in a single piece of prominent malware. Such adware samples are generally much less analyzed than malware. Our revelations call for more concentrated reverse engineering efforts towards adware, and more generally, on PUPs.

\item We track 332 domains used by Wajam to serve injected scripts and updates, and leverage the Umbrella top 1M domain list to estimate Wajam's prevalence over the last two years; we estimate that if Wajam used a single domain, it would rank 5,246th. We also query domains known to be targeted by Wajam through 5M peers from a residential proxy network and find infected peers in 35 countries between 2017 and 2019.

\item We also highlight serious private information leakage and security risks (e.g., enabling MITM with long-lasting effect and possibly RCE attacks) to users affected by Wajam. As new variants remain largely undetected by malware engines during the first days, users even with up-to-date AV/OS remain vulnerable.
\end{enumerate}

\section{Wajam's history}
\label{sec:history}

Wajam Internet Technologies Inc.\ was originally headquartered in Montreal, Canada~\cite{quebec-registry}. Their product (Wajam) aimed at enhancing the search results of a number of websites (e.g., Google, Yahoo, Ask.com, Expedia, Wikipedia, Youtube) with content extracted from a user's social media connections (e.g., Twitter, Facebook, LinkedIn). Wajam was first released in Oct.\ 2011, rebranded as Social2Search in May 2016~\cite{opc-wajam}, then as SearchAwesome in Aug.\ 2017 (as we found).
We use the name Wajam interchangeably to refer to the company or the software they developed.
To gain revenue, Wajam injects ads into browser traffic~\cite{wajam-root1}. 
The company progressively lost its connection with social media and became purely ad/spyware in 2017.\looseness=-1

The OPC Canada investigated the company between Oct.\  2016 and July 2017~\cite{opc-wajam}.
OPC found numerous violations of Canadian Personal Information Protection and Electronic Documents Act (PIPEDA), relative to the egregious collection and preservation of personal data (``\emph{approximately 400 terabytes}'' by the company's own admission), and problematic user consent/EULA, installation/uninstallation methods. 
OPC issued a list of 14 corrective measures. Instead, Wajam sold its activities to a newly created company called Iron Mountain Technology Limited (IMTL) in Hong-Kong, and therefore declared itself unaccountable to Canadian regulations. IMTL seems to have continued Wajam's operations uninterrupted since then and continued to develop its capabilities towards ad injection and AV evasion. We refer the readers interested in the discussion relative to the EULA and user consent to the OPC report.

\section{Related work}
\label{sec:related-work}
Previous studies on worms and botnets mostly focused on the network aspect of such threats, instead of particular software complexity or advanced obfuscation techniques; see e.g., Conficker~\cite{conficker-study}, Torpig~\cite{botnet-takeover} and Mirai~\cite{understand-mirai}. While the largest known botnet reached up to an estimated 50 million users~\cite{storm-botnet}, it is still an order of magnitude lower than the total distribution of Wajam. 

The Mirai botnet was studied across a thousand samples~\cite{understand-mirai}. Authors tracked forks of the original malware, and analyzed the newly added features, including e.g., self-deleting binary, more hardcoded passwords to infect devices---all these changes are largely straightforward.
Moreover, 
Mirai's source code was leaked and readily available. 
In contrast, we reverse-engineer Wajam from scratch to understand the full extent of its capabilities, and bridge significant gaps across generations and major updates, including dealing with e.g., steganography-based installers, custom packers and multiple encryption layers. 

The Zeus banking malware~\cite{zeus-spyeye}, a prominent strain reaching 3.6 million infections, shares some traits with Wajam, including encrypted code sections (albeit done differently), dynamic library loading, encrypted payloads (for configuration files only) with XOR or RC4 hardcoded keys. Zeus also performed MITB by injecting a DLL in browser processes, similar to Wajam's 3rd generation. However, Zeus source code became public in 2016, helping its analysis. Also, active variants of Zeus~\cite{zeus-active} no longer perform browser injection,
in contrast to Wajam's well-maintained browser process injection. \looseness=-1

Targeted Advanced Persistent Threats (APTs) are known for the extent of their operations, both in duration and complexity, e.g.~\cite{stuxnet,apt1-mandian}. In contrast, our focus is an \emph{adware} application, which is not expected to use APT-related techniques e.g., 0-day vulnerabilities. 
Nevertheless, we found that Wajam leverages effective antivirus evasion techniques, and significantly hinders reverse-engineering, over several years. These behaviors are rare even in regular malware. \looseness=-1

Adware can serve as a cover-up for hiding an APT, as it may slip through the hands of an analyst~\cite{carbon-black-apt}. This behavior is coined as Advanced Persistent Adware~\cite{booz-allen-apa}.

Similar to adware, ransomware is also heavily motivated by monetary gains. 
Kharraz et al.~\cite{cutting-gordian-knot} analyzed 1,359 ransomware samples and reported insights into their encryption modules, file replacement and deletion mechanisms. 
Web exploit kits have also been analyzed~\cite{state-exploit-kits,pexy}, including PHP and JavaScript components. The level of sophistication in both cases was limited.

Wajam has been cited in broad analyses covering the distribution models of pay-per-install PUPs~\cite{wajam-usenix-imdea,wajam-usenix-google}; however, only little information about Wajam itself is revealed, including an estimated user base (in the order of $10^7$ during the period Jan.\ 2013--July 2014, much less than the total number of infections reported in the order of $10^8$ by its operators in 2017~\cite{opc-wajam}), and general features (e.g., Wajam is a browser-addon---incorrect since the end of 2014).

In a 2005 report~\cite{symantec2005}, Symantec shows that adware and spyware (without any distinction) exfiltrate sensitive and personally-identifiable data, e.g., extensive system information, names, credit card numbers, username and passwords, or even entire webpages. The use of rootkit techniques, code injection, and random filenames are also discussed.
We not only show that these behaviors are still topical, but we also point at larger security implications resulting from MITM and RCE vulnerabilities, likely due to the lack of incentives from the adware vendor to ship secure code, and from researchers to study and report flaws to such vendors.
Privacy leakages such as browsing histories are also certainly more severe today than they were 14 years ago. In addition, the Internet population, and thus the potential number of victims, has seen a 4-fold increase during this period~\cite{internetworldstats}. Apparently, AV companies used to treat adware more seriously in the past, as evident from the lack of comprehensive reports on recent adware. 

The NetFilter/ProtocolFilters SDKs~\cite{netfilter} were used in PrivDog~\cite{privdog}, which was vulnerable to MITM attacks, as it did not use the certificate validation capabilities of the SDK. B\"{o}ck~\cite{bock-netfilter} extracted the hardcoded private keys from ProtocolFilters found in AdGuard and PrivDog, and listed PUPs that may rely on this library (did not include Wajam).
While PrivDog received significant attention, only one version of the product was vulnerable, affecting 57k users~\cite{privdog}.
The MarketScore spyware also proxied HTTPS traffic~\cite{symantec2005}; however, encrypted traffic was marginal in 2005.
In contrast, Wajam has exposed millions of users to similar MITM attacks for about four years. Compared to Superfish, installed by default on certain Lenovo laptops, Wajam is not bound to a specific hardware vendor. 

Various malicious obfuscation techniques have been documented, including: encrypted code section~\cite{wong2006hunting}, encrypted strings and downloaded configuration files~\cite{DBLP:conf/malware/BlackO16}, junk code~\cite{cisco-obfuscation}, polymorphic icons in Winwebsec, SecurityShield and zbot~\cite{nappa2013driving}, inflated executable file size in the XXMM toolkit~\cite{trick-malware-size}, rootkit as found in the Komodia traffic interception SDK~\cite{komodia-rootkit}, the use of NSIS installers with decryption DLLs in Cerber, Gamarue, Kovter and ZCrypt~\cite{nsis-decryption-malware}, hiding encrypted payloads in BMP~\cite{stegano-bmp} and PNG files~\cite{stegano-png}.
Wajam combines all these techniques from the malware realm, and enhances and layers them. Notably, Wajam's junk code introduces thousands of seemingly purposeful functions interconnected in a dense call graph where the real program functions are hidden. Also, the use of steganography is diversified to various file formats, and is combined with layers of obfuscated encryption and compression in samples from 2018, making Wajam variants highly metamorphic. 

\section{Sample collection and overview}
\label{sec:collection}
We detail below our collection of 52 samples, and summarize their capabilities; 
for their notable features (e.g., the use of code-signing, stealthy installation), see  Table~\ref{tab:wajam-summary} (Appendix).

\subsection{Sample collection}
We obtained our first sample with a known URL to \emph{wajam.com} through the Internet Archive as it is no longer available on the official website.
This sample dates back from Dec.\ 2014, and appears to be a relatively early version of the product.
We obtained 10 more samples from an old malware database~\cite{malekal} by searching for ``Wajam'', two of which were only partial components (DLLs), which we discarded. After analyzing a few samples, we learned about URLs fetched by the application, leading us to query keywords from another malware database~\cite{hybrid-analysis}.
We also learned the URLs serving variants of the installer, and downloaded a sample per month in 2018. At the end of this iterative process, we collected 48 standalone installers, two online installers, and two update packages.

The variants we fetched directly from Wajam servers are named \texttt{Setup.exe}; however, when submitting these samples to VirusTotal, they are sometimes already known by other filenames, e.g., \texttt{update.exe}. We could not find obvious paths that include such filenames on known Wajam servers, suggesting that Wajam is also hosted elsewhere, or downloaded through different vectors. 
As most of the samples are digitally signed and timestamped, or install a signed component, we could trace the history of Wajam over five and a half years, from Jan.\ 2013 to July 2018. 

\subsection{Categories}
\label{sec:categorization}
We identified four injection techniques that were used mostly chronologically. Hence, we refer to each group as a \emph{generation}; see Table~\ref{tab:generations} for the distribution of samples among generations.
We refer to a given sample by its generation letter followed by its chronological index within its generation, e.g., C18. We keep a numerical reference when referring to an entire generation, e.g., third generation.\looseness=-1

\subhead{Generation A: Browser add-on}
The two oldest samples (Jan.\ 2013 and 2014) install add-ons to Chrome, Firefox and IE. There was a Safari add-on as well according to the ``Uninstall'' page on \emph{wajam.com}. A Chrome ad\\
d-on remains available as of Apr.\ 2019, but with only 25 users.
These add-ons were used to directly modify the content of selected websites to insert ads and social-media content in search pages. In samples A1--2, the injection engine, \emph{Priam}, receives search queries and bookmark events.

\subhead{Generation B: FiddlerCore} 
Samples from Sept.\ 2014 to Jan.\ 2016 have their own interception component and leverage the FiddlerCore library~\cite{fiddler-lib} to proxy browser traffic. Each detected browser has its proxy settings set to localhost with a port on which Wajam is listening. HTTPS traffic is broken at the proxy, which certifies the connection by a certificate issued on-the-fly, and signed by a root certificate inserted into the Windows and Firefox trust stores. Only selected domains are intercepted. The application is installed in the Program Files folder with a meaningful name; however, core files have long random names.
Since no component strictly requires a signature by the OS, some samples do not bear any signature. We rely either on a signature on the installer (as seen prior to 2015), or the timestamp of the latest modified file installed (from 2015) to establish a release date for those samples.

\subhead{Generation C: Browser process injection} 
Installers dated between Oct.\ 2014 to May 2016 and two update packages up to Mar.\ 2017 inject a DLL into IE, Firefox and Chrome. In turn, the DLL hooks specific functions to modify page contents after they are fetched from the network (and decrypted in the case of HTTPS traffic), but before they are rendered. Consequently, the injected traffic in encrypted pages is displayed while the browser shows the original server certificate, making this generation more stealthy 
(cf.~\cite{zeus-spyeye,spyeye,citadel-virusbulletin}). 
We tested the latest versions of IE/Firefox/Chrome on an up-to-date Windows 7 32-bit and confirmed that the injection method is still fully functional. We later found that browser hooking parameters are actively maintained and kept updated hourly (Section~\ref{sec:injection-rules}). \looseness=-1

\subhead{Generation D: NetFilter SDK+ProtocolFilters} 
Starting from Apr.\ 2016, a fourth generation implements a NetFilter-based injection technique. Installers dated after May 2016 install a program called Social2Search instead of Wajam. Furthermore, samples dated from Aug.\ 2017 (i.e., few months after the company was sold to IMTL) are again rebranded as SearchAwesome.
The NetFilter SDK enables traffic interception, combined with ProtocolFilters that provides APIs for tampering with the traffic at the application layer.
Instead of explicitly configuring browser proxy settings, NetFilter installs a network driver that intercepts all the network traffic irrespective of the application. In this generation, all HTTPS traffic is intercepted and all TLS connections are broken at the proxy, except for the traffic originating from blacklisted process names. 

\begin{table}[tbp]
	\centering
	\footnotesize
	\caption{Distribution of samples among generations}
	
	\begin{threeparttable}
		\begin{tabular}
			{l|l|r|l}
			\textbf{Gen.} & \textbf{Period covered} & \textbf{\# samples} & \textbf{Injection technique} \\ \hline\hline
			A & 2013-01 -- 2014-07 & 4  & Browser add-on \\ 
			B & 2014-09 -- 2016-01 & 6  & FiddlerCore \\ 
			C & 2014-10 -- 2017-03 & 19 & Browser process injection \\ 
			D & 2016-01 -- 2018-07 & 23 & NetFilter+ProtocolFilters \\ 
		\end{tabular}\par
	\end{threeparttable}
	\label{tab:generations}
\end{table}

\section{Analysis methodology}
\label{sec:methodology}

\subhead{Test environment and sample execution}
We leverage VMware Workstation (WS) and an up-to-date installation of Windows 7 Pro 32-bit with IE 11 and Firefox 61 to capture Wajam's installation process.
For each sample, we instrument WS to start from a fresh VM snapshot, transfer the sample on the guest's desktop, start Process Monitor~\cite{procmon} to capture I/O activities, and start Wireshark on the host OS to record the network traffic. We also take a snapshot of the filesystem and registry before and after the sample is installed to detect modifications made on the system. 

We run the sample with UAC disabled to avoid answering the prompt, and complete the installation, which usually requires clicking only one button at most. It could be possible to instrument the UI to fully automate the process; however, we wanted to verify whether the sample installs without asking for user consent, opens a webpage at the end of the setup, or if the process is completely stealthy.
We note that the UAC prompt is not a significant barrier for Wajam, as it is found bundled (statically or downloaded at runtime) with other installers, for which users already provided admin privileges.\looseness=-1

We could have used existing malware analysis sandboxes; however, a local deployment would have been required as we need control over certain registry keys (e.g., Machine GUID). Furthermore, for consistency and ease of debugging, we used the same environment to capture runtime behaviors and selectively debug samples. \looseness=-1

We also verify the functionality of selected samples on Windows 8.1 Pro 64-bit---some samples lead to a denial of service for certain websites. To fully understand their functionalities, we also conduct a more thorough analysis on selected samples from each generation, by debugging the application and performing MITM attacks.

\subhead{Studying NSIS installers}
Wajam is always based on Nullsoft Scriptable Install System (NSIS~\cite{nsis}), a popular open-source generator of Windows installers~\cite{nsis-dl-stats}. 
NSIS uses LZMA as a preferred compression algorithm and as such, 7-Zip can extract packed files from NSIS-generated installers, unless a modified NSIS is used~\cite{nsis-decompile}. We used 7-Zip for unpacking when possible.
NSIS also compiles an installer based on a configurable installation script written in its own language.
Several NSIS-specific decompilers used to reconstruct the script from installers but trivial modifications in the source code could thwart such automated tools. 7-Zip stopped supporting the decompilation of installer scripts in version 15.06 (Aug.\ 2015)~\cite{7zip-nsis}. We use version 15.05 to successfully decompile these scripts.\looseness=-1

\subhead{Labeling OpenSSL functions}
ProtocolFilters is statically linked with OpenSSL, as indicated by hardcoded strings (e.g., ``RSA part of OpenSSL 1.0.2h~~3 May 2016''). However, IDA FLIRT fails to fingerprint OpenSSL-related functions, even with the help of extra signatures. Given the identified version number, we are able to label essential functions that call \texttt{ERR\_put\_error()}.
Indeed, such calls specify the source file path and line number where an error is thrown, which uniquely identifies a function. By investigating the use of several such functions, we can identify critical sections, e.g., root certificate generation (as used in Section~\ref{sec:flaws}).

\subhead{Debugging}
We leverage IDA Pro and x64dbg~\cite{x64dbg} to debug all binaries to understand some of their anti-analysis techniques.
Due to the extensive use of junk code, identifying meaningful instructions is challenging.
In particular, when reverse-engineering encrypted payloads, we first set breakpoints on relevant Windows API calls to load files (e.g., \texttt{CreateFile}, \texttt{ReadFile}, \texttt{WriteFile}, \texttt{LoadLibrary}), then follow modifications and copies of buffers of interests by setting memory breakpoints on them.
We also rely on interesting network I/O events as seen in Process Monitor to identify relevant functions from the call stack at that time.

To understand the high-level behavior of decryption routines, we combine static analysis and step-by-step debugging.
We also leverage Hex-Rays to study the decompiled code, unless Hex-Rays fails due to obfuscation.
Static analysis is also often made difficult by many dynamic calls resolving only at runtime.

\subhead{Scope}
We focus on reverse-engineering steps that lead to visible consequences on the system and network activities, and document the challenges in doing so. This way, we discover a number of information leaks and several mechanisms to hinder static analysis and evade early antivirus detection. However, we do not claim that we found all such techniques nor that we understand all features of Wajam. Since we do not look at all samples ever released, it is also likely that we missed intermittent features, making our findings a lower bound on Wajam's full potential.

\subhead{Reproducibility}
Since most of this work is a manual effort, we will release intermediate artifacts in an effort to enable reproduction, including: the samples, network traces, file-system and registry modifications during installation, procmon logs, payload decryption scripts, and VT scan logs. The samples include the 52 reverse-engineered ones, the 36 more recent samples scanned with VT, and subsequent samples we kept collecting.

\section{Technical evolution summary}
\label{sec:antianalysis}
We summarize below the inner workings of Wajam and track its changes made over the years---mostly targeted at improving stealthiness and increasing private information leaks. We also demonstrate the efficacy of its evasion techniques by collecting hourly AV detection rates on 36 samples fetched between Aug.\ to Nov.\ 2018.\looseness=-1

\subhead{Wajam modules}
Wajam is composed of several modules, some of which are generation-specific.
Its installer is the first executable an AV gets to analyze, justifying a certain level of obfuscation that constantly increased over time. 
The installer runs a payload (\texttt{brh.dll}, called BRH hereafter) to retrieve system and browser information, e.g., browsing histories, which is then leaked.
The installed binaries comprise the main application, an updater, a browser hooker called ``goblin'' in the 3rd generation, and a persistence module.

\subhead{Typical installation workflow}
A typical sample from 2018 is an NSIS installer with a random icon that unpacks DLLs, which then locate, deobfuscate, decrypt and uncompress a second-stage installer from a media file. In turn, this second installer executes a long obfuscated NSIS script that first calls an unpacked DLL to decrypt and load its BRH companion to perform a number of leaks. Then, it installs the main obfuscated Wajam files under Program Files with random file and folder names. It also adds a persistence module in the Windows directory along with the generated TLS certificate in an `SSL' subdirectory, and a signed network driver (in the \texttt{System32\textbackslash{}drivers} folder). The installer creates three Windows services: 1) the network driver, 2) the main application, 3) the persistence module; and a scheduled task to start the second service at boot time if not already started. 
The main application starts by reading the encrypted updater module, decrypting and executing it. In turn, the module reads the encrypted injection rules, updates them and fetches program updates.

\subhead{Evolution of features}
We provide a timeline with evolution milestones regarding the anti-analysis and evasion techniques,  privacy leaks (more in Section~\ref{sec:antianalysis-leaks}), and new prominent features, in Figure~\ref{fig:timeline}. The timeline also shows the release time of the samples we analyze, labeled on the left when space permits. Techniques are numbered and further discussed in Section~\ref{sec:antianalysis-leaks}.
This evolution illustrates the underlying design of Wajam over the years. In particular, most changes relate to improving the anti-analysis and evasion techniques and could not have been implemented over years had Wajam been stopped by better AV detection. 
Also, between 2014 and early 2017, six types of information leaks were implemented.
For each new feature, the time presented corresponds to the earliest sample we found implementing this feature. Note that all the features do not necessarily accumulate in later samples. For instance, the rootkit capability is found in only three samples.

\newenvironment{timeline}[6]{%
	
	\newcommand{\startyear}{#1}
	\newcommand{\tlendyear}{#2}
	
	\newcommand{\yearcolumnwidth}{#3}
	\newcommand{\rulecolumnwidth}{#4}
	\newcommand{\entrycolumnwidth}{#5}
	\newcommand{\timelineheight}{#6}
	
	\newcommand{\templength}{}
	
	\newcommand{\entrycounter}{0}
	
	\long\def\ifnodedefined##1##2##3{%
		\@ifundefined{pgf@sh@ns@##1}{##3}{##2}%
	}
	
	\newcommand{\ifnodeundefined}[2]{%
		\ifnodedefined{##1}{}{##2}
	}
	
	\newcommand{\drawtimeline}{%
		\draw[timelinerule] (\yearcolumnwidth+5pt, 0pt) -- (\yearcolumnwidth+5pt, -\timelineheight);
		\draw (\yearcolumnwidth+0pt, -10pt) -- (\yearcolumnwidth+10pt, -10pt);
		
		\pgfmathsetlengthmacro{\templength}{neg(add(multiply(subtract(\startyear, \startyear), divide(subtract(\timelineheight, 25), subtract(\tlendyear, \startyear))), 10))}
		\node[year] (year-\startyear) at (\yearcolumnwidth, \templength) {\textbf{\startyear}};
		
	}
	
	\newcommand{\entry}[2]{%
		
		\pgfmathtruncatemacro{\lastentrycount}{\entrycounter}
		\pgfmathtruncatemacro{\entrycounter}{\entrycounter + 1}
		
		\ifdim \lastentrycount pt > 0 pt%
		\node[entry] (entry-\entrycounter) [below of=entry-\lastentrycount] {##2};
		\else%
		\pgfmathsetlengthmacro{\templength}{neg(add(multiply(subtract(\startyear, \startyear), divide(subtract(\timelineheight, 25), subtract(\tlendyear, \startyear))), 10))}
		\node[entry] (entry-\entrycounter) at (\yearcolumnwidth+\rulecolumnwidth+10pt, \templength) {##2};
		\fi
		
		\ifnodeundefined{year-##1}{%
			\pgfmathsetlengthmacro{\templength}{neg(add(multiply(subtract(##1, \startyear), divide(subtract(\timelineheight, 25), subtract(\tlendyear, \startyear))), 10))}
			\draw (\yearcolumnwidth+2.5pt, \templength) -- (\yearcolumnwidth+7.5pt, \templength);
			\node[year] (year-##1) at (\yearcolumnwidth, \templength) {##1};
		}
		
		\draw ($(year-##1.east)+(2.5pt, 0pt)$) -- ($(year-##1.east)+(7.5pt, 0pt)$) -- ($(entry-\entrycounter.west)-(5pt,0)$) -- (entry-\entrycounter.west);
	}
	
	\newcommand{\entrytick}[1]{%
		
		\ifnodeundefined{year-##1}{%
			\pgfmathsetlengthmacro{\templength}{neg(add(multiply(subtract(##1, \startyear), divide(subtract(\timelineheight, 25), subtract(\tlendyear, \startyear))), 10))}
			\node[year] (year-##1) at (\yearcolumnwidth, \templength) {\textbf{##1}};
		}
		
		\draw ($(year-##1.east)+(0pt, 0pt)$) -- ($(year-##1.east)+(10pt, 0pt)$);
	}
	
	\newcommand{\sampletick}[2]{%
		
		\ifnodeundefined{sample-##2}{%
			\pgfmathsetlengthmacro{\templength}{neg(add(multiply(subtract(##1, \startyear), divide(subtract(\timelineheight, 25), subtract(\tlendyear, \startyear))), 10))}
			\node[year] (sample-##2) at (\yearcolumnwidth, \templength) {\textcolor{darkgray}{\tiny##2}};
		}
		
		\draw ($(sample-##2.east)+(2.5pt, 0pt)$) -- ($(sample-##2.east)+(7.5pt, 0pt)$);
	}
	
	\newcommand{\plainentry}[4][1]{
		
		\pgfmathtruncatemacro{\lastentrycount}{\entrycounter}
		\pgfmathtruncatemacro{\entrycounter}{\entrycounter + 1}
		
		\ifdim \lastentrycount pt > 0 pt%
		\node[entry] (entry-\entrycounter) [below of=entry-\lastentrycount, yshift=neg((##1-1) * 1.5mm)] {##4};
		\else%
		\pgfmathsetlengthmacro{\templength}{neg(add(multiply(subtract(\startyear, \startyear), divide(subtract(\timelineheight, 25), subtract(\tlendyear, \startyear))), 10))}
		\node[entry] (entry-\entrycounter) at (\yearcolumnwidth+\rulecolumnwidth+10pt, \templength) {##4};
		\fi
		
		\ifnodeundefined{invisible-year-##3}{%
			\pgfmathsetlengthmacro{\templength}{neg(add(multiply(subtract(##2, \startyear), divide(subtract(\timelineheight, 25), subtract(\tlendyear, \startyear))), 10))}
			\draw (\yearcolumnwidth+2.5pt, \templength) -- (\yearcolumnwidth+7.5pt, \templength);
			\node[year] (invisible-year-##3) at (\yearcolumnwidth, \templength) {};
		}
		
		\draw ($(invisible-year-##3.east)+(2.5pt, 0pt)$) -- ($(invisible-year-##3.east)+(7.5pt, 0pt)$) -- ($(entry-\entrycounter.west)-(5pt,0)$) -- (entry-\entrycounter.west);
	}
	
	\begin{tikzpicture}
	\tikzstyle{entry} = [%
	align=left,%
	text width=\entrycolumnwidth,%
	node distance=3mm,%
	anchor=west]
	\tikzstyle{year} = [anchor=east]
	\tikzstyle{timelinerule} = [%
	draw,%
	decoration={markings, mark=at position 1 with {\arrow[scale=1.5]{latex'}}},%
	postaction={decorate},%
	shorten >=0.4pt]
	
	\drawtimeline
}
{
	\end{tikzpicture}
}
\makeatother

\newcommand{\tPolymorphism}{1\xspace}
\newcommand{\tIcons}{2\xspace}
\newcommand{\tNestExec}{3\xspace}
\newcommand{\tPayloadEnc}{4\xspace}
\newcommand{\tStego}{5\xspace}
\newcommand{\tCustomEnc}{6\xspace}
\newcommand{\tObfsKey}{7\xspace}
\newcommand{\tObfsNSIS}{8\xspace}
\newcommand{\tObfsNetPS}{9\xspace}
\newcommand{\tWhitelist}{10\xspace}
\newcommand{\tMRT}{11\xspace}
\newcommand{\tInflated}{12\xspace}
\newcommand{\tObfsString}{13\xspace}
\newcommand{\tDynamicCalls}{14\xspace}
\newcommand{\tJunkCode}{15\xspace}
\newcommand{\tEncCode}{16\xspace}
\newcommand{\tIdaTricks}{17\xspace}
\newcommand{\tStringArgs}{18\xspace}
\newcommand{\tRandomNames}{19\xspace}
\newcommand{\tRootkit}{20\xspace}
\newcommand{\tPersistence}{21\xspace}
\newcommand{\tDetection}{22\xspace}
\newcommand{\tSignatures}{23\xspace}

\begin{figure}[t]
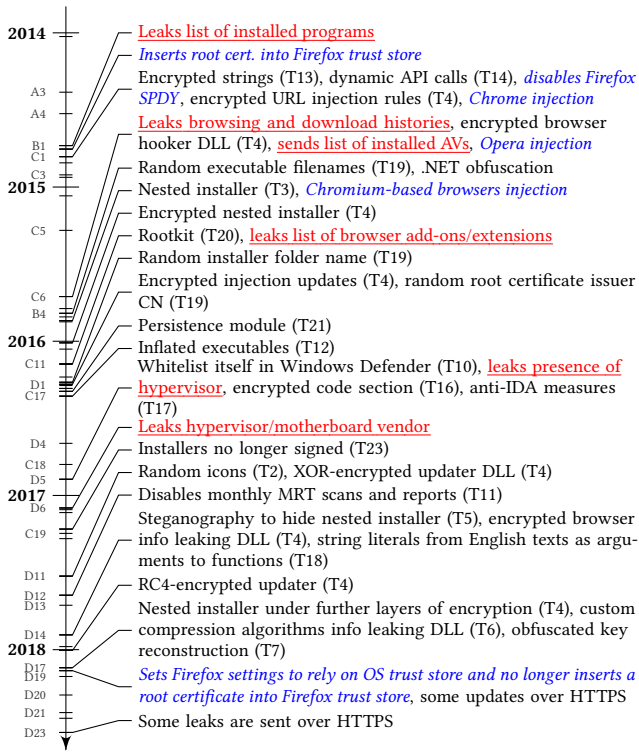

	\footnotesize
	\begin{timeline}{2014}{2018.4}{1.25cm}{.7cm}{6.6cm}{3.9in}
		\plainentry{2014.731}{B1}{\textcolor{red}{\ul{Leaks list of installed programs}}}
		\plainentry{2014.754}{B2}{\textcolor{blue}{\emph{Inserts root cert.\ into Firefox trust store}}}
		\plainentry[2]{2014.803}{C1}{Encrypted strings (T\tObfsString), dynamic API calls (T\tDynamicCalls), \textcolor{blue}{\emph{disables Firefox SPDY}}, encrypted URL injection rules (T\tPayloadEnc), \textcolor{blue}{\emph{Chrome injection}}}
		\plainentry[3]{2015.710}{C6}{\textcolor{red}{\ul{Leaks browsing and download histories}}, encrypted browser hooker DLL (T\tPayloadEnc), \textcolor{red}{\ul{sends list of installed AVs}}, \textcolor{blue}{\emph{Opera injection}}}
		\plainentry[2]{2015.819}{B4}{Random executable filenames (T\tRandomNames), .NET obfuscation}
		\plainentry{2015.866}{C8}{Nested installer (T\tNestExec), \textcolor{blue}{\emph{Chromium-based browsers injection}}}
		\plainentry{2016.010}{C10}{Encrypted nested installer (T\tPayloadEnc)}
		
		\plainentry{2016.147}{C11}{Rootkit (T\tRootkit), \textcolor{red}{\ul{leaks list of browser add-ons/extensions}}}
		\plainentry{2016.266}{C14}{Random installer folder name (T\tRandomNames)}
		\plainentry[2]{2016.279}{D1}{Encrypted injection updates (T\tPayloadEnc), random root certificate issuer CN (T\tRandomNames)}
		\plainentry[2]{2016.325}{D3}{Persistence module (T\tPersistence)}
		\plainentry{2016.356}{C17}{Inflated executables (T\tInflated)}
		\plainentry[2.5]{2016.895}{D5}{Whitelist itself in Windows Defender (T\tWhitelist), \textcolor{red}{\ul{leaks presence of hypervisor}}, encrypted code section (T\tEncCode), anti-IDA measures (T\tIdaTricks)}
		\plainentry[2.5]{2017.081}{D6}{\textcolor{red}{\ul{Leaks hypervisor/motherboard vendor}}}
		\plainentry{2017.218}{D9}{Installers no longer signed (T\tSignatures)}
		\plainentry{2017.523}{D11}{Random icons (T\tIcons), XOR-encrypted updater DLL (T\tPayloadEnc)}
		\plainentry{2017.647}{D12}{Disables monthly MRT scans and reports (T\tMRT)}
		\plainentry[3]{2017.904}{D14}{Steganography to hide nested installer (T\tStego), encrypted browser info leaking DLL (T\tPayloadEnc), string literals from English texts as arguments to functions (T\tStringArgs)}
		\plainentry[3]{2018.005}{D16}{RC4-encrypted updater (T\tPayloadEnc)}
		\plainentry[3]{2018.117}{D17}{Nested installer under further layers of encryption (T\tPayloadEnc), custom compression algorithms info leaking DLL (T\tCustomEnc), obfuscated key reconstruction (T\tObfsKey)}
		\plainentry[4]{2018.136}{D18}{\textcolor{blue}{\emph{Sets Firefox settings to rely on OS trust store and no longer inserts a root certificate into Firefox trust store}}, some updates over HTTPS}
		\plainentry[2]{2018.537}{D23}{Some leaks are sent over HTTPS}
		
		\entrytick{2015}
		\entrytick{2016}
		\entrytick{2017}
		\entrytick{2018}
		
		\sampletick{2014.024}{}
		\sampletick{2014.385}{A3}
		\sampletick{2014.524}{A4}
		\sampletick{2014.731}{B1}
		\sampletick{2014.754}{}
		\sampletick{2014.803}{C1}
		\sampletick{2014.843}{}
		\sampletick{2014.921}{C3}
		\sampletick{2014.937}{}
		\sampletick{2015.057}{}
		\sampletick{2015.281}{C5}
		\sampletick{2015.710}{C6}
		\sampletick{2015.788}{}
		\sampletick{2015.819}{B4}
		\sampletick{2015.842}{}
		\sampletick{2015.866}{}
		\sampletick{2015.874}{}
		\sampletick{2016.010}{}
		\sampletick{2016.051}{}
		\sampletick{2016.147}{C11}
		\sampletick{2016.150}{}
		\sampletick{2016.232}{}
		\sampletick{2016.266}{}
		\sampletick{2016.279}{D1}
		\sampletick{2016.287}{}
		\sampletick{2016.306}{}
		\sampletick{2016.306}{}
		\sampletick{2016.325}{}
		\sampletick{2016.356}{C17}
		\sampletick{2016.662}{D4}
		\sampletick{2016.799}{C18}
		\sampletick{2016.895}{D5}
		\sampletick{2017.081}{D6}
		\sampletick{2017.092}{}
		\sampletick{2017.111}{}
		\sampletick{2017.218}{}
		\sampletick{2017.246}{C19}
		\sampletick{2017.280}{}
		\sampletick{2017.523}{D11}
		\sampletick{2017.647}{D12}
		\sampletick{2017.712}{D13}
		\sampletick{2017.904}{D14}
		\sampletick{2017.980}{}
		\sampletick{2018.005}{}
		\sampletick{2018.117}{D17}
		\sampletick{2018.136}{}
		\sampletick{2018.174}{D19}
		\sampletick{2018.294}{D20}
		\sampletick{2018.408}{D21}
		\sampletick{2018.445}{}
		\sampletick{2018.537}{D23}
	\end{timeline}
	\vspace{-.15in}
	\caption{Timeline of first appearance of key features (colors: black $\to$ anti-analysis/evasion improvements, \textcolor{blue}{\emph{blue}} $\to$ new functional features, \textcolor{red}{\ul{red}} $\to$ information leaks)}
	\label{fig:timeline}
\end{figure}

\begin{figure}[t]
	\centering
	\includegraphics[width=\columnwidth]{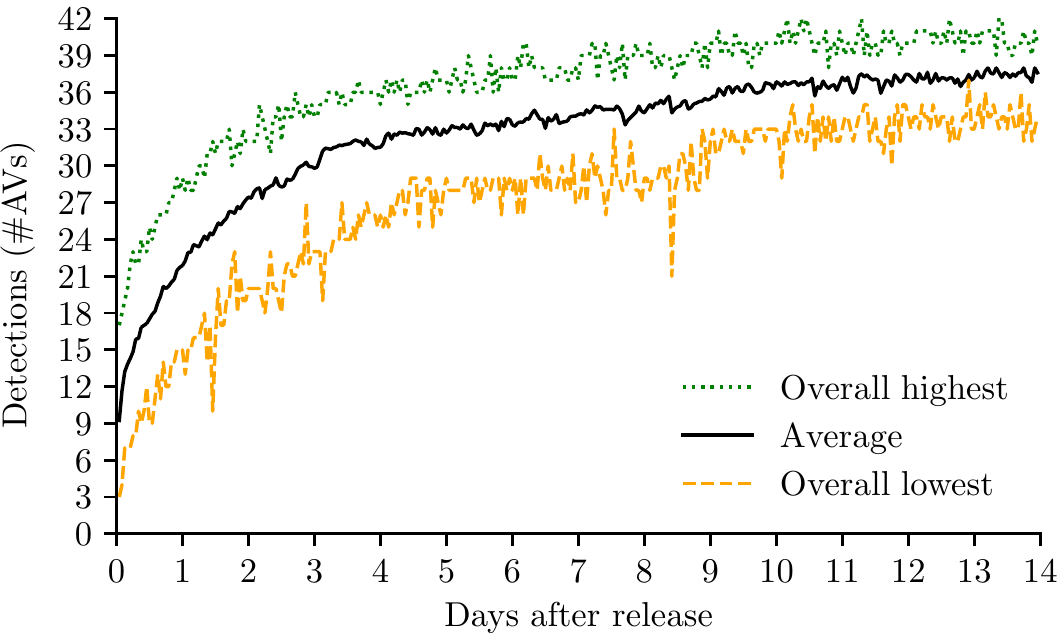}
	\vspace{-.15in}
	\caption{VirusTotal detection rates of 36 samples starting from their release time}
	\label{fig:av-rates}
\end{figure}

\subhead{Antivirus detection rates}
We submitted samples to VirusTotal that we obtained directly from one of Wajam's servers. We pooled a known URL to retrieve daily samples as soon as possible after they are released to observe early detection rates. In total, we collected 36 samples between Aug.---Nov.\ 2018; see Fig.~\ref{fig:av-rates} for the VirusTotal detection rates. The rates are given relative to the release time as indicated by the ``Last-Modified'' HTTP header provided by the server. We trigger a rescan on VirusTotal approximately every hour after the first submission to observe the evolution for up to two weeks. \looseness=-1

Fig.~\ref{fig:av-rates} illustrates the averaged rates, along with the overall lowest and highest rates during each hour. The rates converge to about 37 detections out of about 69 AV engines at the end of the two-week period. 
Note that the total number of AV engines slightly changes over time, as reported by VT.
Importantly, we notice that the rates start arguably low during the first hours. The lowest detection ratio of 3/68 is found on the Aug.\ 8 sample, 19min after its release. Only one AV labels Wajam correctly, another one identifies it as different malware, and the third one simply labels it ``ML.Attribute.HighConfidence.''
Similarly, the sample from Apr.\ 29 is flagged by 4/71 AVs, three of them label it with ``heuristic'', ``suspicious'' and ``Trojan.Generic,'' suggesting that they merely detect some oddities.
The average rate during the first hour is only about 9 AVs. The quick rise in the number of detections in the first 2--3 days are hindered by new daily releases that restart the cycle from a low detection rate. We believe this strategy has helped Wajam continue to spread for years despite the (late) detections.

Moreover, Wajam is rarely labeled as is by AVs. Rather, they often output generic names\footnote{
	``\emph{Win32.Adware-gen}'', 
	``\emph{heuristic}'', 
	``\emph{Trojan.Gen.2}'', 
	``\emph{Unsafe}''} 
or mislabel samples.\footnote{
	``\emph{Adware.Zdengo}'', 
	``\emph{Gen:Variant.Nemesis.430}''} 
Certain AVs label Wajam as PUP/not-a-virus/Riskware/Optional;\footnote{
	``\emph{Generic PUA PC (PUA)}'', 
	``\emph{PUP/Win32.Agent.C2840632}'', 
	``\emph{PUA:Win32/Wajam}'', 
	``\emph{not-a-virus:HEUR:AdWare.Win32.Agent.gen}'', 
	``\emph{Pua.Wajam}'', 
	``\emph{Riskware.NSISmod!}'', 
	``\emph{Riskware}'', 
	``\emph{PUP.Optional.Wajam}''} 
however, we note that depending on the configuration of such AVs, no alert or action may be triggered upon detection, or the alert may show differently than for regular malware~\cite{install-top10,giri2016alerting}. Also, once installed, the detection rate of the installer is irrelevant. Rather, the detection of individual critical files matter.
For instance, while D23's detection rate is 35/66 AVs after 15 days, its installed files remain less detected: 26/66 for the uninstaller after 16 days, 16/67 for the main binary after 22 days, and 9/69 for the network driver after 26 days.

\section{Prevalence}
\label{sec:prevalence}

We illustrate the prevalence of Wajam through the popularity of its domains and a brief overview of  worldwide infections.

\subsection{Domains popularity}

First, we list the domain names used by Wajam, as found in code signing certificates, hardcoded URLs in samples, ad injection rules we downloaded, and domains declared in legal documents of the company~\cite{quebec-registry}.
We also gather domain names that were hosted simultaneously from the same IP address or subnet,\footnote{We leverage historical DNS data from \emph{DnsTrails.com}.} then manually verify whether they resemble other Wajam domains.
We also rely on domains found in CT logs that follow the pattern \emph{technologie*.com} or \emph{*technology.com}, as we found it is recurrent. We query all the 14,944 matching domains and keep the ones that serve a webpage referring to Wajam/Social2Search/SearchAwesome (similar to \emph{wajam.com}), share the same favicon as previously identified, or distribute Wajam's installer from a predefined URL. The complete list of 332 domains is provided in Table~\ref{tab:wajam-domains} (Appendix). Note that not all Wajam domains may follow these patterns, thus our domain list is a lower bound on the total number of domains used.

This domain list rarely evolves over time, and most domains follow the common pattern mentioned above. During our study, they were hosted in France (OVH, under the same /24 subnet) and the US (Secured Servers). Some served browser-trusted certificates issued by RapidSSL until Mar.\ 2018, then by Let's Encrypt. Many domains were never issued a certificate.

\begin{figure}[t]
	\centering
	\includegraphics[width=\columnwidth]{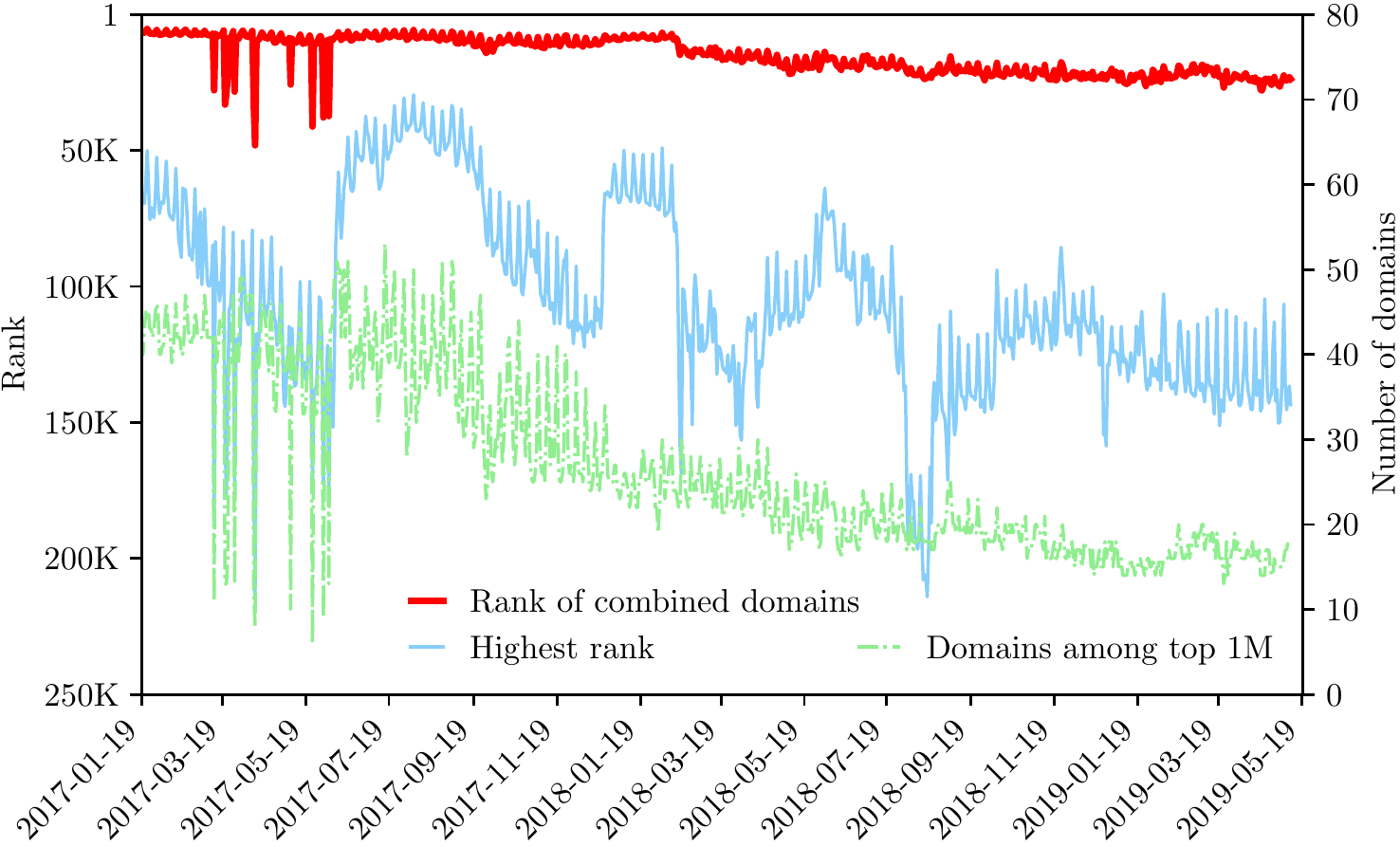}
	\vspace{-.1in}
	\caption{Wajam domains in Umbrella's top list (2017--2019)}
	\label{fig:wajam-domains}
\end{figure}

We then search for the rank of these domains in Umbrella's top 1M domain list from 2017 to 2019. Umbrella is the only public list that tracks domain popularity from DNS queries, and thus, captures the popularity of domains pooled for updates by Wajam, as well as those serving ads after injection. Fig.~\ref{fig:wajam-domains} shows the number of Wajam domains per daily list along with the highest ranking of these domains.
Over the last two years, we found as many as 53 domains with the top ranked one reaching the 29,427th position.

However, given the number of domains concurrently used, the highest rank is not the best measure to represent the overall domains popularity.
Borrowing the idea from Le Pochat et al.~\cite{LePochat2019}, we consider that the popularity follows a Zipf distribution and combine all Wajam domains into one rank by following the Dowdall rule. This rule attributes a weight to each domain that is inversely proportional to its rank. The rank of a combination of domains is the inverse of the sum of their weights.
If all Wajam domain requests were intended to only one domain, this domain would rank between 27,895th and 5,246th during the past 28 months (ignoring the sudden drops in the first half of 2017). Such a rank indirectly hints at a significant number of infections.

We note a slight decline in popularity over this period; however, it may not necessarily correlate with a reduction of Wajam's activities, i.e., the popularity is only relative. Also, our domain list may miss newer popular domains, especially if they do not follow the identified naming scheme.\looseness=-1

\subsection{Worldwide infections}
We leverage a residential proxy service (Luminati\footnote{Advertised with 40M peers, \url{https://luminati.io}}) to query 89 domains where Wajam injects ads.
Each peer runs a client that allows other peers (i.e., us) to relay network traffic through it.
We found that Wajam only relies on a blacklist of processes (see Section~\ref{sec:flaws}), which does not include the Luminati client process name. Therefore, if Wajam has infected a peer, we expect that our traffic will be intercepted by the peer's Wajam instance, and we should obtain a Wajam-issued certificate for the domains queried.

We consider the domains found in the 101 injection rules fetched in Jan.\ 2019 (see Section~\ref{sec:flaws}), then we remove Google- and LinkedIn-related domains since Luminati does not permit querying them. We then establish TLS connections to these domains through 4.2M peers in all countries available.
Note that the domains only relate to search engines and shopping websites, thus no illegitimate or dangerous websites are accessed through the peers. In addition, due to high bandwidth costs of Luminati, we only establish a TLS connection and retrieve the certificate, then close the socket, i.e., no HTTP query is made. Using this setup, we can only detect the second and fourth generations. Since the third generation only modifies traffic by hooking selected browser processes, a Luminati peer infected with this generation would not intercept our traffic. 

To detect Wajam-issued certificates, we rely on fingerprints we established based on the reverse-engineering of the certificate generation (see Section~\ref{sec:directions-detection}).
We performed our scans in Mar.\ 2019. 
We detected 52 cases in 25 countries: Indonesia (10 infected peers), Malaysia (4); Argentina, India, Italy, Philippines (3); Brazil, Canada, Chile, France, Honduras, Spain, Thailand, Vietnam (2); Australia, C\^ote d'Ivoire, Colombia, Denmark, Ecuador, Mexico, Netherlands, Peru, Russia, the US, and Venezuela (1).

During a similar scan we conducted through Luminati in June 2017 through 911k peers in only 33 countries from Reporter Without Border's list~\cite{rwb-list}, we detected 214 cases in 19 countries: Vietnam (98 infected peers), India (42), Malaysia (16), Thailand (12), the UK (7), Hong Kong (6), Belarus (5), Venezuela (5); Egypt, France, Libya, Pakistan (3); Iran, Russia, Turkey, the US (2); South Korea, Sri Lanka, and Yemen (1).

Note that peers on Luminati network are not necessarily representative of the general population, therefore the proportion of infections might not be informative. However, Wajam was found in a total of 35 countries between 2017 and 2019, highlighting the scope of its infections.

\section{Private information leaks}
\label{sec:leaks}
Beyond installing the files onto the system, the installer also performs other core tasks, including the generation of unique IDs, and leaking browsing and download histories. We detect these leaks from the network captures and trace their origin into the binaries to better understand them.

Two unique identifiers are generated during installation based on a combination of the MAC address, user folder path, and disk serial number. These IDs are appended to all requests made to Wajam's servers and ads distributors. They are used for ad tracking, and to detect repeated installations to identify pay-per-install frauds by Wajam distributors, i.e., a distributor faking numerous installations to increase its revenue from Wajam~\cite{opc-wajam}.

From B1, the installer leaks the list of installed programs as found in the registry, minus Microsoft-specific updates in some cases.
The OS version and the date of the installation obtained from Wajam's own timestamping service, are also sent in each query.

From C6, the browsing history of IE, Firefox and Chrome is sent in plaintext to Wajam's servers, along with the history of Opera from D6. Only the newest sample we analyzed, dated from July 2018, sends this information over HTTPS.
This leak is the most privacy-sensitive. For users who do not configure an expiration of their history, the leak could span over several months' worth of private data.
In Chrome, the local history expires after three months~\cite{chrome-history}, mitigating the extent of the leak; however, other browsers do not expire their history, which could last for years.
In parallel, the download history, i.e., the URLs of downloaded files, is also sent in plaintext except in the latest sample.

After the installation, Wajam continues to send the list of browser addons/extensions, installed programs, and detected AVs whenever it fetches updates from the server.

Samples dated after the end of 2016 (from D5) check whether they are running on a virtual machine  by calling the CPUID instruction. The result is appended to all HTTP(S) queries made by the installer, along with the BIOS manufacturer name, which could also expose the hypervisor.
We are unsure about the consequences of this reporting as we still observed fully functional samples in our VMs (with complete updates and injected ads).

\section{Anti-analysis and evasion}
\label{sec:antianalysis-leaks}

Wajam leverages at least 23 techniques to hinder static analysis, fingerprinting, reverse engineering, and antivirus detection:
\tPolymorphism) metamorphism, \tIcons) changing static resources, \tNestExec) nested executables, \tPayloadEnc) payload compression and encryption, \tStego) steganography, \tCustomEnc) custom encryption and encoding, \tObfsKey) obfuscated key reconstruction, \tObfsNSIS) obfuscated installer script, \tObfsNetPS obfuscated .NET and PowerShell, \tWhitelist) auto-whitelisting in Windows Defender, \tMRT) disabling MRT, \tInflated) inflated files, \tObfsString) string obfuscation and encryption, \tDynamicCalls) dynamic API calls, \tJunkCode) junk and dead code, \tEncCode) encrypted code, \tIdaTricks) anti-IDA Pro measures, \tStringArgs) unique readable strings as function arguments, \tRandomNames) randomized names, \tRootkit) rootkit, \tPersistence) persistence/resurrection module, \tDetection) detection of installed antiviruses (only leaks the result), and \tSignatures) digital signatures (or the lack thereof).

We discuss below the newer techniques and those that have been improved or are specific to Wajam; for others, see Appendix~\ref{sec:antianalysis-details}.

\subhead{T\tPolymorphism: Metamorphism}
The main technique is to produce metamorphic variants, i.e., an obfuscated packer that changes dynamically its logic around the same template and evolves through generations. It unpacks varying payloads that perform similar actions. This translates in numerous variants, which are released daily, mostly around 3--5pm UTC since at least 2018. Variants seems to be released automatically, hence it would be interesting to identify the underlying generator. However, we could not find any name or fingerprint.

\subhead{T\tStego: Steganography}
Starting from D14, the installer unpacks a handful of small DLL files, and a large picture or audio file (MP3, WAV, BMP, GIF, PNG).
At first, this media file appears to contain only random audio noise or colors, and could be a simple dummy file only useful to arbitrarily inflate the installer's size (cf.~\cite{trick-malware-size}). 
The DLLs are, in fact, used to reconstruct an encrypted compressed nested installer. The payload is simply stuffed into data sections of the media file.
For instance, in D14, an MP3 file is composed of MPEG frames starting with a four-byte header 
and followed by 622 bytes of data. We found that the DLL extracts and concatenates the data section from each frame to reconstruct a GZip file, which in turn reveals a second NSIS installer.
From D20, the payload starts from an arbitrary offset, complicating automated deobfuscation.

To the best of our knowledge, only few cases of malware leveraging steganography are known, and they relied on a single format and trivial encryption~\cite{stegano-bmp,stegano-png}. Wajam thus brings steganography to multiple formats, with added obfuscation.

\begin{table*}[th]
	\centering
	\small
	\setlength{\tabcolsep}{5pt}
	\caption{Steganographic techniques to hide a nested installer in samples from end-2017 to 2018}
	\vspace{-.1in}
	\begin{threeparttable}
		\begin{tabular}
			{l|l|l|p{2.2in}|l}
			\textbf{ID} & \textbf{Hidden in} & \textbf{Payload reconstruction} & \textbf{Encryption/Compression} & \textbf{Stream encryption keys} \\ \hline \hline
			D14--15 & MP3 & Concatenated MPEG frame data & plaintext (GZip) & \emph{Not applicable} \\
			D16 & MP3 & Concatenated MPEG frame data & custom encryption & \emph{Not applicable} \\
			D17 & GIF & In section after LSD + custom offset & custom stream cipher+compression & \texttt{2njZEYFf}, \texttt{qsjmoRZ7FM} \\
			D18 & BMP & BitmapLine section + custom offset & custom stream cipher+encryption+compression & \texttt{ldXTyqwQ}, \texttt{ckXKI19jmC}\\
			D19 & WAV & First DataChunk samples + custom offset & custom stream cipher+compression & \texttt{47txnKuG}, \texttt{eyimwKIOBG} \\
		\end{tabular}\par
	\end{threeparttable}
	\label{tab:stegano}
\end{table*}

\subhead{T\tCustomEnc: Custom encryption and encoding}
While payload encryption was usually done with RC4 or XOR, a custom stream cipher is used starting from D17 for the nested installer, outlined in Algorithm~\ref{fig:xor}. From D20, the encryption becomes difficult to comprehend as it involves more than 2000 decompiled lines of C code, with numerous branches and inter-dependent loops. The decryption seems to update an intermediate state, and may likely be a stream cipher; however, we could not identify which one. Alternatively, it could be a form of encoding since we could not find an associated key either.
Malware is known to modify encryption routines; however, the changes are small enough and the underlying algorithm is still identifiable, e.g., modified RC4 in Citadel~\cite{DBLP:conf/malware/BlackO16}.

\begin{algorithm}[t]
	\footnotesize
	\renewcommand{\algorithmicrequire}{\textbf{Input:}}
	\renewcommand{\algorithmicensure}{\textbf{Output:}}
	\caption{Custom stream cipher in samples D17 and above}
	\label{fig:xor}
	\begin{algorithmic}
		\Require ciphertext $c$, first key $key_1$, second key $key_2$
		\Ensure plaintext $p$
		\State $p \gets []$
		\For{$i$ from $0$ to $\text{len}(c)-1$}
		\State $p[i] \gets c[i] \oplus key_1[i \mod \text{len}(key_1)]$
		\State $key_1[i] \gets p[i]$
		\State $p[i] \gets p[i] \oplus key_2[i \mod \text{len}(key_2)]$
		\State $key_2[i] \gets p[i]$
		\EndFor
	\end{algorithmic}
\end{algorithm}

\subhead{T\tInflated: Inflated files}
Some malware scanners are known to discard large files~\cite{bullguard-filesize,clamav-conf}, hence an obvious anti-analysis technique is to inflate the size of the executable.
Seven samples rely on enlarged \texttt{.rdata} (C17, D4) or code sections (D6--10), resulting in binaries ranging from 9 to 26MiB in size.
The first type consists of a large \texttt{.rdata} section that contains strings duplicated hundreds of times. However, this section contains actual strings used in the unobfuscated application. Given that such strings are meant to be decrypted at runtime, it is unclear why the developers left plaintext strings in the binary, or if large \texttt{.rdata} sections are at all meant for evasion. 
Large code sections tend to slow IDA Pro's analysis, possibly due to gibberish instructions parsed.

The goblin DLL is also sometimes decrypted at runtime and written back to disk, at which point it is inflated by appending 10MiB of apparently random data.
In addition, the size of the installer increases over time and heavily fluctuates in the fourth generation, between 4--10MiB, depending on the size of the installed files. In turn, the unpacked file sizes depend on T\tJunkCode.

\subhead{T\tJunkCode: Junk and dead code}
Junk/dead code usually involves adding, replacing, or reordering instructions~\cite{rad2012camouflage,gao2014survey}.
Wajam's junk code is quite distinct from what can be found in the literature. It involves: 1) string manipulation on large random strings, 2) inter-register operations, 3) calls to Windows library functions that only swap or return some fixed values, 4) tests on the result of such dummy functions, 5) large never-executed conditional branches, and 6) dependence on global variables.
Useful operations are thus interleaved with such junk code.
Due to modifications that are sometimes made to global variables common to many functions, these functions are not deterministic from their inputs, thus junk code removal is challenging.
For instance, in D17, the DLLs that read and decode media files (T\tStego), contain more than 2000 and 400 junk functions, respectively, that can be called up to a dozen times each. The resulting call graph is also useless.

\subhead{T\tIdaTricks: Anti-IDA Pro measures}
Encrypted code (T\tEncCode) involves multi-MiB placeholders in the code section to receive decrypted instructions (the decryption is not in-place). They are pre-filled with a single byte padding. As a byproduct of this technique, both the padding and encrypted instructions are difficult to analyze by a disassembler.
For instance, IDA Pro hangs for over two hours on sample D9, containing 4MiB of the byte \texttt{B9} (interpreted as a jump instruction), followed by another 3MiB of encrypted instructions.

\subhead{T\tStringArgs: Unique readable strings as function arguments}
Often, functions are called with an argument that is a unique random string, or a brief extract from public texts; e.g., we found strings from the Polish version of \emph{Romeo and Juliet} in D14--16, and from \textit{The Art of War} by Sun Tzu in D17,19,23. This technique could be used to thwart heuristics (based on entropy or human-readable text); however, we are unsure about its intended target.

\subhead{T\tSignatures: Digital signatures}
Before D9, samples are digitally signed, which could help the installer appear legitimate to users when prompted for administrative rights (when distributed as a standalone app), and lower detection by AVs~\cite{certified-pup}. From D9 (i.e., shortly after Wajam was sold to IMTL), only the network drivers are still signed, as required by Windows. Presumably, since the signing certificates are issued to Wajam's domains, which could help AVs to fingerprint the installer, and hence signatures were removed. Also, Wajam already inherits admin privileges from the bundled software installer that runs it and no longer triggers Windows UAC prompts. From D20, the main installed binaries are also signed.

\section{Security threats}
\label{sec:flaws}
In this section, we discuss the security flaws we identified in Wajam's TLS proxy certificate validation, along with vulnerabilities of its auto-update mechanisms that lead to arbitrary content injection (with possible persistence) and privileged remote code execution.

\subhead{Certificate validation issues}
In the 2nd and 4th generations, Wajam acts as a TLS proxy, and therefore is expected to validate server certificates.
FiddlerCore-based samples (2nd gen.) properly do so.
However, in ProtocolFilters-based samples (4th gen.), Wajam fails to validate the hostname, since at least Apr.\ 2016 (D1). Thus, a valid certificate for \emph{example.com} is accepted by Wajam for \emph{any} other domain. Worse, Wajam even replaces the Common Name (CN) from the server certificate with the domain requested by the client. In turn, the browser accepts the certificate for the requested domain as it trusts Wajam's root certificate. 

Swapping the CN with the requested domain is somewhat mitigated, since 1) CAs should include a Subject Alternate Name (SAN) extension in their certificates, which is copied from the original certificate by ProtocolFilters, and 2) browsers may ignore the CN field in a certificate if a SAN extension is present. In particular, Chrome rejects certificates without SAN~\cite{chrome-san}. Consequently, if an attacker obtains a valid certificate for any domain without a SAN extension, they are still able to perform a MITM attack against IE and Firefox when Wajam is installed. 

Despite the deprecation of CN as a way of binding a certificate to a domain~\cite{cn-deprecated} in 2000, Kumar et al.~\cite{cert-misissuance} recently showed that one of the most common errors in certificate issuance by publicly trusted CAs is the lack of a SAN extension.
For the sake of our experiment, we inserted our own root certificate in the Windows trust store and issued a certificate without SAN for \emph{evil.com}. Wajam successfully accepted it when visiting \emph{google.com}, and the Wajam-generated certificate in turn was accepted by IE.

\subhead{Shared root private key}
We located the code in ProtocolFilters responsible to create the root certificate used for interception.
The code either generates an RSA-2048 private key (using OpenSSL), or use a default hardcoded one. Unfortunately, the default settings are used and all 4th generation samples share the same key.
We performed a successful MITM attack on our test system using a test domain. 
Consequently, an attacker could impersonate any HTTPS websites to a machine running Wajam's fourth generation by knowing the root certificate's CN to properly chain the generated certificates.
However, the CN is based on the Machine GUID, as illustrated in Table~\ref{tab:root-certs} (more details in Appendix~\ref{sec:cn-generation}).

\begin{table}[t]
	\centering
	\footnotesize
	\setlength{\tabcolsep}{4pt}
	\caption{TLS root certificates in 2nd and 4th generations}
	
	\begin{threeparttable}
		\begin{tabular}
			{l|l}
			\textbf{Sample} & \textbf{Root certificate's Common Name} \\
			\hline
			
			B1--B3  & Wajam\_root\_cer              \\
			B4--B5  & WNetEnhancer\_root\_cer       \\
			B6      & WaNetworkEnhancer\_root\_cer  \\
			D1--D2  & \texttt{md5(GUID+`WajaInterEn')[0:16]} \\
			D3      & \texttt{md5(GUID+`WNEn')[0:16]}          \\
			D4      & \texttt{md5(GUID+`Social2Se')[0:16]}     \\
			D5--D8  & \texttt{md5(GUID+`Socia2Sear')[0:16]}    \\
			D9      & \texttt{md5(GUID+`Socia2Se')[0:16]}      \\
			D10     & \texttt{md5(GUID+`Socia2S')[0:16]}       \\
			D11     & \texttt{md5(GUID+`Soci2Sear')[0:16]+` 2'}     \\
			D12--D21& \texttt{md5(GUID+`SrcAAAesom')[0:16]+` 2'}    \\
			D22--D23& \texttt{base64(md5(GUID+`SrcAAAesom')[0:12])+` 2'}    \\
		\end{tabular}\par
	\end{threeparttable}
	\label{tab:root-certs}
	
\end{table}

Since the Machine GUID is unpredictable and generally unknown to an attacker, and since the resulting CN carries at least 48 bits of entropy in our dataset (starting from D22, 64 bits in prior samples), crafting certificates signed by a target Wajam's root certificate is generally impractical. Indeed, an attacker would need to serve an expected number of $2^{47}$ certificates to a victim before one is accepted. We note that environments with cloned Windows installations across hosts could be more vulnerable if the Machine GUID is not properly regenerated on each host, as it is possible to obtain it from a single host with few privileges.

Nevertheless, during our scans through residential proxies (see Section~\ref{sec:prevalence}), we also found cases of injected scripts pointing to Wajam domains with much shorter issuer CNs, e.g., ``MDM5Z 2'' providing under 15 bits of entropy (see Appendix~\ref{sec:cn-generation}). This could indicate more recent variants are at higher risks of MITM attacks. 

The FiddlerCore-based generation is immune to this issue as keys are randomly generated at install-time using \emph{MakeCert}.

\subhead{Auto-update mechanism}
Wajam periodically fetches traffic injections rules, browser hooking configurations, and program updates. Updates are fetched upon first launch, then Wajam waits for a duration indicated in the last update (from 50 to 80 minutes in our tests), before it updates again.
While early samples fetched plaintext files, all recent samples and the whole 4th generation download encrypted files.
The decryption is handled in an encrypted DLL loaded at runtime. We found that Wajam uses the MCrypt library to decrypt updates with a hardcoded key and IV using the Rijndael-256 cipher (256-bit block, key and IV) in CFB-8 mode. 
The key and IV are the same across all versions. The content of such updates and the implications of lacking the proper protection are discussed below. \looseness=-1

\subhead{Downgraded website security}
From the 2nd generation, Wajam fetches traffic injections rules, containing a list of domains and instructions to inject scripts.
The injection file is a JSON structure containing ``supported websites.''
For each website, a list of regular expressions are provided to match URLs of interest, often specifically about search or item description pages, along with specific JavaScript and CSS URLs to be injected from one of Wajam's several possible domains. The rules also include HTTP headers or tags to be added or removed.

Since the content injection relies on loading a remote third-party script, browsers may refuse to load the content due to mixed-content policies, or the Content Security Policy (CSP) configured by the website. Mixed-content is addressed by loading the script over the same protocol as the current website. For websites that specify a CSP HTTP header or HTML tag, Wajam removes this CSP from the server's response before it reaches the browser, to ensure their script is properly loaded.
Wajam removes the CSP from Facebook, \emph{mail.ru}, Yandex, \emph{flipkart.com}, and Yahoo Search; see Fig.~\ref{fig:inj-facebook} where the CSP header is  dropped from \emph{facebook.com}.

Other response headers are also removed in some cases, including \texttt{Access-Control-Allow-Origin}, which would allow the given website's resources to be fetched from different origins than those explicitly allowed by the website, and \texttt{X-Frame-Options} (e.g., on \emph{rambler.ru}), enabling the website to be loaded in a frame.

Such behaviors not only allow injected scripts to be successfully loaded and fetch information, but also effectively downgrade website security (e.g., XSS vulnerabilities may become exploitable).

\begin{figure}[t]
  \centering
  \footnotesize
  \begin{verbatim}
[facebook]
  [domains]
    [0] => facebook
  [patterns]
    [0] =>
     ^https?:\/\/(www\.)?facebook.com(?!(\/xti\.php))
  [js]
    [0] =>
     se_js.php?se=facebook&integration=searchenginev2
  [css]
  [headers]
    [remove]
      [response]
        [0] => content-security-policy
  \end{verbatim}
  \caption{Example of traffic injection rule for \texttt{facebook.com} that matches all pages except \texttt{xti.php}}
  \label{fig:inj-facebook}
  
\end{figure}

\subhead{Arbitrary content injection} 
Traffic injection rules are always fetched over plain HTTP.
Although updates are encrypted, an attacker can learn the encryption algorithm and extract the hardcoded key/IV from any Wajam sample in the last few years, to easily forge updates and serve them to a victim through a simple MITM attack.

\begin{figure}[t]
	\footnotesize
	\vspace{-1em}
\begin{lstlisting}[style=mylisting,language=json]
{"version":"1",
 "update_interval":60,
 "base_url":"\/\/attacker.evil\/",
 "supported_sites":
   {"bank":
     {"domains":["bank"],
      "patterns":["^https?:\\\/\\\/login\\.bank\\.com"],
      "js":["bank.js"],
      "css":[],"version":"1"}},
 "process_blacklist":[],
 "process_whitelist":[],
 "update_url":"https:\/\/attacker.evil\/mapping",
 "css_base_url":"\/\/attacker.evil\/css\/",
 "url_filtering":[],
 "bi_events":[],
 "url_tracking":[],
 "protocols_support":
   {"quic_udp_block":1}}
\end{lstlisting}
	\vspace{-.1in}
	\caption[Traffic injection rule to insert a malicious script on \url{login.bank.com}]{Traffic injection rule to insert a malicious script on \url{login.bank.com} located at \url{//attacker.evil/bank.js}, and redirect future update queries to \url{https://attacker.evil/mapping}}
	
	\label{fig:bank-inject}
	
\end{figure}

As a proof-of-concept, we suppose that \emph{bank.com} is a banking website with its login page at \url{https://login.bank.com}. We craft an update file that instructs Wajam to insert a JavaScript file of our choice, hosted on our own server, and encrypt it using the key that we recovered. The plaintext traffic injection rule is provided in Fig.~\ref{fig:bank-inject}.
Once the update is fetched by Wajam (i.e., after around an hour, or at boot time), and upon visiting the bank's login page, our malicious script is loaded on the bank's page and is able to manipulate the page's objects, including listening to keystroke events in the username and password fields.
No default cross-origin policy would prevent our attack. If the bank's website implemented a CSP, it could be easily removed from the server's HTTP response.

We note that Wajam already has the infrastructure in place for maliciously injecting \emph{any} script into \emph{any} website at will, by simply distributing malicious updates. Such updates could be short-lived for stealthiness, yet affect a large number of victims.

Moreover, updates systematically contain the URL of the next update to fetch.
Once Wajam downloads an update and caches it to disk, it does not use its hardcoded URL anymore.
Hence, the effect of a compromised update is persistent.
Our malicious update (Fig.~\ref{fig:bank-inject}) instructs Wajam to fetch further updates from our own server, alleviating the need to repeatedly perform MITM attacks.

\subhead{Privileged remote code execution}
Wajam also queries for program updates and retrieves the manifest of potential new versions.
Several parameters are passed, including Wajam's current version, and the list of detected security solutions, possibly influencing which update is served.
If an update is available, the URL where to fetch a ZIP package is provided, which is downloaded and uncompressed into the installation directory.

Similar to the attack on traffic injection rules, it is possible to serve a fake update manifest to trigger an update from a malicious URL before mid-Feb.\ 2018 (D18), while software updates were fetched over HTTP. This would enable an attacker to inject its own binary that will be run with SYSTEM privileges; however, we have not tested this attack. Starting from D18, software updates are fetched over HTTPS and it appears that Wajam properly validates the server certificate, mitigating this attack.

\section{Content injection}
We discuss below the domains targeted for injection, and the content injected into webpages. We also summarize the specificities of the 3rd generation that conducts MITB attacks.

\subsection{Targeted domains}
The injection rules fetched between Feb.\ to July 2018 always include 100 regular expressions to match the domains of major websites, with only one change during this period. The injected domains include popular search engines, social networks, blogging platforms, and various other localized businesses in North America, Western Europe, Russia, and Asia. The list contains notable websites, e.g., Google, Yahoo, Bing, TripAdvisor, eBay, BestBuy, Ask, YouTube, Twitter, Facebook, Kijiji, Reddit, as well as country-specific websites, e.g., \emph{rakuten.co.jp}, \emph{alibaba.com}, \emph{baidu.com}, \emph{leboncoin.fr}, \emph{willhaben.at}, \emph{mail.ru}.
The total number of websites that are subject to content injection is not easy to quantify due to the nature of some URL matching rules, e.g., in the case of the blogging platform Wordpress, blogs are hosted as a subdomain of \emph{wordpress.com} and Wajam's rules match \emph{any} subdomain, which could be several millions~\cite{wordpress-stats}.

\begin{figure*}[t]
	\footnotesize
	\begin{lstlisting}[style=mylisting,language=HTML5]
	<script data-type="injected" src="//technologietravassac.com/addon/script/google?integration=searchenginev2&har=2&v=n11.14.1.86&os_mj=6&os_mn=1&os_bitness=32&mid=b8230ac083f9fb5067a66e03b4882491&uid=B77FCD732C2E5337FF907BFAA44758D1&aid=3673&aid2=none&ts=1531782569&ts2="></script>
	<link rel="stylesheet" type="text/css" href="//main-social2search.netdna-ssl.com/css/cdn/min_search_engine_v2.css?wv=1.00434"/>
	\end{lstlisting}
	\vspace{-.2in}
	\caption{Example of injected content on \url{google.com}}
	\label{fig:wajam-inserted-script}
	
\end{figure*}

\subsection{Injected content}
\label{sec:injected-content}
On URLs matching the injection rules, Wajam injects a JavaScript and CSS right before the \texttt{</head>} tag, a feature provided by ProtocolFilters. The scripts are either self-contained in early samples, or they insert remote scripts with parameters including Wajam's version, the OS version/architecture, the two unique IDs (see Section~\ref{sec:leaks}), an advertiser ID, and the installation timestamp; see Fig.~\ref{fig:wajam-inserted-script}. 
The remote JavaScript URL script injected into the page is dependent on the visited website. Two categories of websites are distinguished here: search engines, and shopping websites.
We give below an example for each case.

\subhead{Search engines}
There are three possible behaviors that we observed when visiting a search engine website.
For instance, when searching on \emph{google.com}, Wajam can change the action on the first few results' links returned by Google. In effect, when a user clicks on these results, the original link opens in a new browser tab while the original tab loads a series of ad trackers (including Yahoo and Bing) provided with the keywords searched by the user, and eventually lands on an undesirable page, e.g., a search result page from \emph{informationvine.com} about foreign exchange.
Alternatively, the script may just redirect the user to \emph{searchpage.com}, a domain that belongs to Wajam, which in turn redirects to a Yahoo search result page with the user's original search keywords. A user may not notice that her original search on Google is eventually served by Yahoo. In the meantime, her keyword searches are sent to Wajam's server. Also, the Yahoo result URL contains parameters that may indicate an affiliation with Wajam, i.e.,
\texttt{hspart=wajam} and \texttt{type=wjsearchpage\_ya\_3673\_fja6rh1}.
Finally, Wajam may simply insert several search results that it fetched from its servers, as the top results. Wajam performs a seamless integration of those results in the page, breaching the trust that a user has in the search engine results. This behavior is part of a patent owned by Wajam Internet Technologies Inc~\cite{archambault2013method}.

\subhead{Shopping websites}
When searching on \emph{ebay.com}, Wajam loads a 180KiB JavaScript file (more than 7700 SLOC) containing the Priam engine intended to retrieve search keywords, fetch related ads, and integrate them on the page. This engine seems self-contained and embeds several libraries. It has numerous methods to manipulate page elements and cookies. Inserted ads are shown at the top of the result list in a large format, also seamlessly integrated, thanks to injected CSS. When the user clicks one of the ads, she is redirected to a third party website selling products related to her search.

In both cases, one of the unique IDs generated by Wajam's installer accompanies each URL pointing to Wajam's domains. In the end, both Wajam and the advertisers can build a profile of the user based on her searches.

\subsection{Browser hooking rules}
\label{sec:injection-rules}
The third generation specifically retrieves a browser hooking configuration file with offsets of functions to be hooked in a number of browsers and versions.
Unlike the traffic injection rules, the browser hooking rules are preloaded in the installer. Hence, it is possible to study their evolution in time.

The earliest third generation sample (Nov.\ 2014, C1) only includes addresses of functions to be hooked for 47 versions of Chrome, from version 18 to 39. The file also lists supported versions of IE and Firefox, although old and without specific function addresses.
In Sept.\ 2015 (C6), Wajam introduces the support for seven versions of the Opera browser. Two months later, five other Chromium-based browsers are introduced, of which four are adware, i.e., BrowserAir, BoBrowser, CrossBrowser, MyBrowser; and one is a legitimate browser intended for  Vietnamese users, i.e., Coc Coc.
By Jan.\ 2016 (C10), 200 versions of Chrome are supported, up to version 49.0.2610.0 with finer granularity for intermediate versions.\looseness=-1

Wajam's browser hooking DLL name was blacklisted in Chrome in Nov.\ 2014~\cite{chrome-blacklist-wajam} because it could cause crashes. Other blacklisted DLLs are labeled in the comments as adware, malware or keylogger, but Wajam is not. One month later (in C3), Wajam randomized this DLL name, making the blacklist ineffective.

Although we did not capture any new sample from the third generation after Jan.\ 2016, we noticed that the browser hooking rules are kept up-to-date, suggesting that this generation is still actively maintained and possibly distributed. In an update from July 2018, we count 1176 supported Chrome versions including the latest Canary build, and additional Chromium-based browsers, e.g., Torch, UC Browser, and Amigo Browser. Versions of Opera are outdated by more than a year. Other Chromium-based browsers only have entries for a limited number of selected versions.

Wajam avoids intercepting non-browser applications as evident from a blacklist of process names in the update file, e.g., dropbox.exe, skype.exe, bittorrent.exe. Additionally, a whitelist is also present, including the name of supported browser processes; however, it appears not to be used.
Furthermore, Wajam seems to have had difficulties handling certain protocols and compression algorithms in the past. It disables SPDY in Firefox and SDCH compression in Chrome before v46.

\section{Directions for better detection}
\label{sec:directions-detection}
Security solutions overall fail to statically analyze Wajam's installers and binaries. Unless such binaries are submitted for analysis, possibly because they look suspicious and endpoint solutions may decide to upload them to the antivirus cloud, Wajam can still be installed on most user systems due to its daily metamorphic installer.
We identified simple fingerprints that could hint at an infection, either from the host or network activities.
First, Wajam registers an installed product on the system using either a known registry key or known names (e.g., SearchAwesome), which could be blacklisted.
Then, it tries to add its installation folder and network driver as exceptions for Windows Defender, which could help locate Wajam's binaries.
Moreover, Wajam uses a long but bounded list of domains so far. A simple domain blacklist would prevent Wajam to communicate with its servers and leak private information.
Samples communicating in plaintext can further be fingerprinted due to the URL patterns and type of data sent, i.e., list of installed programs.
Later samples that leverage HTTPS at install-time and later to fetch updates could still be fingerprinted due to known domains present in the TLS SNI extension, or simply by blacklisting corresponding IP addresses.
Since daily variants of Wajam are served from known domains at known directories, it is possible for security solutions to constantly monitor these servers for new samples and create corresponding signatures earlier.
When a new system driver is installed, additional verifications could quickly find out Wajam's driver as it is signed with a certificate for one of the known domains.

Finally, we were able to build fingerprints for Wajam-issued certificates, shown in Table~\ref{tab:wajam-tls-fingerprints}. It is possible to match a leaf certificate's distinguished name (DN) with our patterns to confirm whether it has been issued by Wajam. They may be particularly relevant if integrated into browsers to warn users. Chrome already detects well-known software performing MITM to alert users of possible misconfigurations or unwanted interceptions~\cite{chrome-mitm-list}.

The use of ProtocolFilters can also be fingerprinted by the files and folder structure it sets up. Online searches for \emph{malware ``2.cer''} and \emph{``SSL'' ``cert.db'' ``*.cer''} yield several forum discussions about infections, e.g., Win.Dropper.Mikey, iTranslator, ContentProtector, SearchProtectToolbar, GSafe, OtherSearch, and even a security solution (Protegent Total Security, from India). Most of these applications likely use ProtocolFilters' default key, as we could verify for Protegent, and hence make end users vulnerable to MITM attacks, in addition to being a nuisance. More work is needed to understand the extent of the use of this interception SDK.

\section{Wajam clones} 
While searching for other ProtocolFilters-enabled applications, we also stumbled upon OtherSearch (also known as FlowSurf/CleverAdds). This adware application shares very similar obfuscation, evasion and steganography techniques with Wajam, sometimes in a more or less advanced way, to the point that it is mislabeled as Wajam when detected by AVs.
For instance, it disables MRT (T\tMRT) and also SmartScreen, and randomizes file paths as done in Wajam (T\tRandomNames). The installer also leverages steganography (T\tStego) to run a second installer hidden in media files; however, it uses a custom ZIP extractor instead of NSIS.
Moreover, OtherSearch also embeds ProtocolFilters' default key in its root certificate, but does not randomize the issuer names (T\tRandomNames), thus exposing all its victims to trivial MITM attacks on HTTPS traffic.
However, OtherSearch does not leak the browser histories.
We did not observe variants served daily at known URLs, thus we are unsure whether OtherSearch leverages such poly/metamorphism technique (T\tPolymorphism).

We could not find an organizational connection between Wajam and OtherSearch, thus suggesting that both may leverage a common third-party obfuscation framework, or simply share similar ideas.
A recent report by McAfee suggests that adware vendors delegate the obfuscation job to ``freelancers''~\cite{mcafee-wakenet}. Hence, the same third party could have been hired by both businesses.

We also note that one network request, made during the installation of OtherSearch to report a successful installation, triggers a non-interpreted PHP script on the server side; this leaks the credentials for an Internet-facing MySQL database. We gathered simple statistics over this database and found that it contains over 100 million Google searches and associated clicked results from the past 1.5 years (nearly 20GiB). 6.54M records are associated with unique IDs, indicating a large number of potential victims.
Two third of the searches seem to originate from France, as hinted by the domain \emph{google.fr} in the search queries.
We reported the whereabouts of this database to the hosting provider (OVH) and on the French Ministry of Interior's report platform on Apr.\ 17, 2019.

\section{Concluding remarks}
\label{sec:conclusion}

Apparently, the adware business is a Pandora's Box that stayed overlooked for too long, which leverages interesting known and newer anti-analysis techniques for successful evasion, and results into disastrous security and privacy violations. If such threats were taken seriously, the bar could easily be raised to thwart the most ludicrous of them. For instance, the 332 domains that belong to Wajam could be tracked and blacklisted. The daily released samples issued from some of these domains could be monitored and blacklisted within minutes. Fixed registry keys created during installation that have not changed in years are enough to kill all related processes and quarantine them. Unfortunately, this is not the case as of today.

Compared to previous recent studies on adware, we provide an in-depth look into a widespread strain in particular, and provide insights into the business and technical evolutions. We uncovered several anti-analysis and antivirus evasion techniques. We also identified important security risks and privacy leakages. Considering the huge amount of private data collected by its operators, and the number of installations it made, it is surprising that it remained virtually overlooked and fully functional for many years. Perhaps, ``adware'' applications do not present themselves as much attractive targets for analysis. However, we hope that the security community will recognize the need for better scrutiny of such applications, and more generally PUPs, as they tend to survive and evolve into more robust variants. 

\section*{Acknowledgments}
This work is party supported by a grant from CIRA.ca's Community Investment Program. 
The first author was supported in part by a Vanier Canada Graduate Scholarship (CGS).
The second author is supported in part by an NSERC Discovery Grant.


\appendix

\begin{table*}[!t]
	\centering
	\fontsize{6.5}{7.4}\selectfont
	\caption{Samples summary (N/A means not applicable, e.g., expired downloader samples do not install an application)}
	\vspace{-.05in}
	\setlength{\tabcolsep}{4.6pt}
	\begin{threeparttable}
		\begin{tabular}
			{l|p{1.3in}|>{\centering\arraybackslash}m{.2in}|l|l|l|c|c|c|c|l}
			\multicolumn{1}{l}{\textbf{ID}} & \multicolumn{1}{l}{\begin{minipage}{1.3in}\textbf{Installer/downloader/\newline patch filename}\end{minipage}} & \mcrot{1}{l}{45}{\begin{minipage}{.6in}\textbf{Signed\newline component?}\end{minipage}} & \multicolumn{1}{l}{\textbf{Date UTC}} & \multicolumn{1}{l}{\textbf{Authenticode CN}} & \multicolumn{1}{l}{\textbf{Installed name}} & \mcrot{1}{l}{45}{\textbf{Autoinstall}} & \mcrot{1}{l}{45}{\textbf{Opens webpage}} & \mcrot{1}{l}{45}{\textbf{Stealthy}} & \mcrot{1}{l}{45}{\textbf{Rootkit}} & \multicolumn{1}{l}{\textbf{Origin}} \\
			\hline
			\hline
			A1 & 
			wajam\_install.exe & 
			\yes & 
			2013-01-03 & 
			Wajam & 
			Wajam & 
			~ & 
			\yes & 
			~ & 
			~ & 
			Hybrid Analysis \\
			
			\hline
			A2 & 
			wajam\_setup.exe & 
			\yes & 
			2014-01-09 & 
			Wajam Internet Technologies Inc & 
			Wajam & 
			~ & 
			~ & 
			~ & 
			~ & 
			Hybrid Analysis \\
			
			A3 & 
			wajam\_download.exe & 
			\yes & 
			2014-05-21 & 
			Insta-Download.com & 
			\textit{N/A} & 
			\textit{N/A} & 
			\textit{N/A} & 
			\textit{N/A} & 
			~ & 
			Malekal MalwareDB \\
			
			A4 & 
			wajam\_download\_v2.exe & 
			\yes & 
			2014-07-11 & 
			Insta-Download.com & 
			\textit{N/A} & 
			\textit{N/A} & 
			\textit{N/A} & 
			\textit{N/A} & 
			~ & 
			Malekal MalwareDB \\
			
			B1 & 
			WIE\_2.15.2.5.exe & 
			\yes & 
			2014-09-25 & 
			FastFreeInstall.com & 
			Wajam & 
			~ & 
			\yes & 
			~ & 
			~ & 
			Malekal MalwareDB \\
			
			B2 & 
			WIE\_2.16.1.90.exe & 
			\yes & 
			2014-10-03 & 
			FastFreeInstall.com & 
			Wajam & 
			~ & 
			\yes & 
			~ & 
			~ & 
			Malekal MalwareDB \\
			
			C1 & 
			WWE\_1.1.0.48.exe & 
			\yes & 
			2014-10-21 & 
			AutoDownload.net & 
			Wajam & 
			~ & 
			\yes & 
			~ & 
			~ & 
			VirusShare \\
			
			C2 & 
			WWE\_1.1.0.51.exe & 
			\yes & 
			2014-11-05 & 
			AutoDownload.net & 
			Wajam & 
			~ & 
			\yes & 
			~ & 
			~ & 
			VirusShare \\
			
			C3 & 
			WWE\_1.2.0.31.exe & 
			\yes & 
			2014-12-03 & 
			AutoDownload.net & 
			Wajam & 
			~ & 
			\yes & 
			~ & 
			~ & 
			VirusShare \\
			
			B3 & 
			wajam\_setup.exe & 
			\yes & 
			2014-12-09 & 
			Wajam Internet Technologies Inc & 
			Wajam & 
			~ & 
			\yes & 
			~ & 
			~ & 
			Archive.org \\
			
			\hline
			C4 & 
			WWE\_1.2.0.53.exe & 
			\yes & 
			2015-01-21 & 
			AutoDownload.net & 
			Wajam & 
			~ & 
			\yes & 
			~ & 
			~ & 
			VirusShare \\
			
			C5 & 
			wwe\_1.43.5.6.exe & 
			\yes & 
			2015-04-13 & 
			installation-sur-iphone.com & 
			Wajam & 
			~ & 
			\yes & 
			~ & 
			~ & 
			Hybrid Analysis \\
			
			C6 & 
			WWE\_1.52.5.3.exe & 
			\yes & 
			2015-09-17 & 
			chabaneltechnology.com & 
			Wajam & 
			\yes & 
			\yes & 
			~ & 
			~ & 
			Hybrid Analysis \\
			
			C7 & 
			WWE\_1.53.5.19.exe & 
			\yes & 
			2015-10-16 & 
			trudeautechnology.com & 
			Wajam & 
			\yes & 
			\yes & 
			~ & 
			~ & 
			Hybrid Analysis \\
			
			B4 & 
			WIE\_2.38.2.13.exe & 
			~ & 
			2015-10-27 & 
			\textit{N/A} & 
			Wajam & 
			~ & 
			\yes & 
			~ & 
			~ & 
			Malekal MalwareDB \\
			
			B5 & 
			wie\_2.39.2.11.exe & 
			~ & 
			2015-11-05 & 
			\textit{N/A} & 
			Wajam & 
			~ & 
			\yes & 
			~ & 
			~ & 
			Malekal MalwareDB \\
			
			C8 & 
			wajam\_install.exe & 
			\yes & 
			2015-11-13 & 
			preverttechnology.com & 
			Wajam & 
			\yes & 
			\yes & 
			~ & 
			~ & 
			Malekal MalwareDB \\
			
			C9 & 
			WWE\_1.55.1.20.exe & 
			\yes & 
			2015-11-16 & 
			preverttechnology.com & 
			Wajam & 
			\yes & 
			\yes & 
			~ & 
			~ & 
			Hybrid Analysis \\
			
			\hline
			C10 & 
			WWE\_1.58.101.25.exe & 
			\yes & 
			2016-01-04 & 
			yvonlheureuxtechnology.com & 
			Wajam & 
			\yes & 
			\yes & 
			~ & 
			~ & 
			Hybrid Analysis \\
			
			B6 & 
			WIE\_2.40.10.5.exe & 
			~ & 
			2016-01-19 & 
			\textit{N/A} & 
			Wajam & 
			\yes & 
			~ & 
			\yes & 
			~ & 
			Hybrid Analysis \\
			
			C11 & 
			WWE\_1.61.80.6.exe & 
			\yes & 
			2016-02-23 & 
			saintdominiquetechnology.com & 
			\textit{(nothing)} & 
			\yes & 
			\yes & 
			~ & 
			\yes & 
			Hybrid Analysis \\
			
			C12 & 
			WWE\_1.61.80.8.exe & 
			\yes & 
			2016-02-24 & 
			saintdominiquetechnology.com & 
			Wajam & 
			\yes & 
			\yes & 
			~ & 
			~ & 
			Hybrid Analysis \\
			
			C13 & 
			WWE\_1.63.101.27.exe & 
			\yes & 
			2016-03-25 & 
			carmenbienvenuetechnology.com & 
			Wajam & 
			\yes & 
			\yes & 
			~ & 
			~ & 
			Hybrid Analysis \\
			
			C14 & 
			WWE\_1.64.105.3.exe & 
			\yes & 
			2016-04-07 & 
			Telecharger-Installer.com & 
			Wajam & 
			\yes & 
			\yes & 
			~ & 
			~ & 
			Hybrid Analysis \\
			
			D1 & 
			WBE\_0.1.156.12.exe & 
			\yes & 
			2016-04-11 & 
			technologieadrienprovencher.com & 
			Wajam & 
			\yes & 
			\yes & 
			~ & 
			~ & 
			VirusShare \\
			
			C15 & 
			WWE\_1.65.101.8.exe & 
			\yes & 
			2016-04-14 & 
			sirwilfridlauriertechnology.com & 
			Wajam & 
			\yes & 
			\yes & 
			~ & 
			~ & 
			VirusShare \\
			
			D2 & 
			wbe\_0.1.156.16.exe & 
			\yes & 
			2016-04-21 & 
			technologieadrienprovencher.com & 
			Wajam & 
			\yes & 
			\yes & 
			~ & 
			~ & 
			VirusShare \\
			
			C16 & 
			WWE\_1.65.101.21.exe & 
			\yes & 
			2016-04-21 & 
			sirwilfridlauriertechnology.com & 
			Wajam & 
			\yes & 
			\yes & 
			~ & 
			~ & 
			VirusShare \\
			
			D3 & 
			WBE\_3.5.101.4.exe & 
			\yes & 
			2016-04-28 & 
			technologieadrienprovencher.com & 
			Wajam & 
			\yes & 
			\yes & 
			~ & 
			~ & 
			Hybrid Analysis \\
			
			C17 & 
			wwe\_9.66.101.9.exe & 
			\yes & 
			2016-05-09 & 
			sirwilfridlauriertechnology.com & 
			Social2Search & 
			\yes & 
			\yes & 
			~ & 
			\yes & 
			VirusShare \\
			
			D4 & 
			WBE\_11.8.1.26.exe & 
			\yes & 
			2016-08-29 & 
			technologieferronnerie.com & 
			Social2Search & 
			\yes & 
			\yes & 
			~ & 
			~ & 
			Hybrid Analysis \\
			
			C18 & 
			patch\_1.68.15.18.zip & 
			\yes & 
			2016-10-18 & 
			beaubourgtechnology.com & 
			\textit{N/A} & 
			\textit{N/A} & 
			\textit{N/A} & 
			\textit{N/A} & 
			\yes & 
			wajam-download.com \\
			
			D5 & 
			WBE\_crypted\_bundle\_11.12.1.100 .release.exe & 
			\yes & 
			2016-11-22 & 
			emersontechnology.com & 
			Social2Search & 
			\yes & 
			\yes & 
			~ & 
			~ & 
			Hybrid Analysis \\
			
			\hline
			D6 & 
			WBE\_crypted\_bundle\_11.12.1.301 .release.exe & 
			\yes & 
			2017-01-30 & 
			wottontechnology.com & 
			Social2Search & 
			\yes & 
			\yes & 
			~ & 
			~ & 
			Malekal MalwareDB \\
			
			D7 & 
			WBE\_crypted\_bundle\_11.12.1.310 .release.exe & 
			\yes & 
			2017-02-03 & 
			piddingtontechnology.com & 
			Social2Search & 
			\yes & 
			\yes & 
			~ & 
			~ & 
			Hybrid Analysis \\
			
			D8 & 
			WBE\_crypted\_bundle\_11.12.1.334 .release.exe & 
			\yes & 
			2017-02-10 & 
			quaintontechnology.com & 
			Social2Search & 
			\yes & 
			\yes & 
			~ & 
			~ & 
			Hybrid Analysis \\
			
			D9 & 
			WBE\_crypted\_bundle\_11.13.1.52 .release.exe & 
			\yes & 
			2017-03-21 & 
			wendleburytechnology.com & 
			Social2Search & 
			\yes & 
			\yes & 
			~ & 
			~ & 
			Hybrid Analysis \\
			
			C19 & 
			patch\_1.77.10.1.zip & 
			~ & 
			2017-04-01 & 
			\textit{N/A} & 
			\textit{N/A} & 
			\textit{N/A} & 
			\textit{N/A} & 
			\textit{N/A} & 
			~ & 
			wajam-download.com \\
			
			D10 & 
			WBE\_crypted\_bundle\_11.13.1.88 .release.exe & 
			\yes & 
			2017-04-13 & 
			technologieflagstick.com & 
			Social2Search & 
			\yes & 
			\yes & 
			~ & 
			~ & 
			Hybrid Analysis \\
			
			D11 & 
			Setup.exe & 
			\yes & 
			2017-07-11 & 
			terussetechnology.com & 
			Social2Search & 
			\yes & 
			~ & 
			~ & 
			~ & 
			Hybrid Analysis \\
			
			D12 & 
			Setup.exe & 
			\yes & 
			2017-08-25 & 
			vanoisetechnology.com & 
			SearchAwesome & 
			\yes & 
			~ & 
			~ & 
			~ & 
			Hybrid Analysis \\
			
			D13 & 
			Setup.exe & 
			\yes & 
			2017-09-18 & 
			technologievanoise.com & 
			SearchAwesome & 
			\yes & 
			~ & 
			~ & 
			~ & 
			Hybrid Analysis \\
			
			D14 & 
			s2s\_install.exe & 
			\yes & 
			2017-11-27 & 
			boisseleautechnology.com & 
			SearchAwesome & 
			\yes & 
			~ & 
			~ & 
			~ & 
			Hybrid Analysis \\
			
			D15 & 
			update.exe & 
			\yes & 
			2017-12-25 & 
			barachoistechnology.com & 
			SearchAwesome & 
			\yes & 
			~ & 
			~ & 
			~ & 
			Hybrid Analysis \\
			
			\hline
			D16 & 
			Setup.exe & 
			\yes & 
			2018-01-02 & 
			technologienouaillac.com & 
			SearchAwesome & 
			\yes & 
			~ & 
			~ & 
			~ & 
			Hybrid Analysis \\
			
			D17 & 
			Setup.exe & 
			\yes & 
			2018-02-12 & 
			pillactechnology.com & 
			SearchAwesome & 
			\yes & 
			~ & 
			~ & 
			~ & 
			Hybrid Analysis \\
			
			D18 & 
			Setup.exe & 
			\yes & 
			2018-02-19 & 
			pillactechnology.com & 
			SearchAwesome & 
			\yes & 
			~ & 
			~ & 
			~ & 
			Hybrid Analysis \\
			
			D19 & 
			Setup.exe & 
			\yes & 
			2018-03-05 & 
			technologiepillac.com & 
			SearchAwesome & 
			\yes & 
			~ & 
			~ & 
			~ & 
			mileendsoft.com \\
			
			D20 & 
			Setup.exe & 
			\yes & 
			2018-04-18 & 
			monestiertechnology.com & 
			SearchAwesome & 
			\yes & 
			~ & 
			~ & 
			~ & 
			technologiesnowdon.com \\
			
			D21 & 
			Setup.exe & 
			\yes & 
			2018-05-30 & 
			bombarderietechnology.com & 
			SearchAwesome & 
			\yes & 
			~ & 
			~ & 
			~ & 
			technologiesnowdon.com \\
			
			D22 & 
			Setup.exe & 
			\yes & 
			2018-06-12 & 
			technologiebombarderie.com & 
			SearchAwesome & 
			\yes & 
			~ & 
			~ & 
			~ & 
			technologiesnowdon.com \\
			
			D23 & 
			Setup.exe & 
			\yes & 
			2018-07-16 & 
			technologievouillon.com & 
			SearchAwesome & 
			\yes & 
			~ & 
			~ & 
			~ & 
			technologiesnowdon.com \\
			\hline
		\end{tabular}\par
		\footnotesize
		\textbf{Legend:}
		The ``Filename'' is the most descriptive name we found from either the source where we found the sample, HA~\cite{hybrid-analysis} or VirusTotal.
		``Signed component'' indicates whether the installer or a component it installs is authenticode-signed, in which case the Date column refers to the authenticode signature date, otherwise it shows the latest file timestamp among all installed files.
		``Authenticode CN'' reflects the corresponding Common Name on the signing certificate.
		``Installed name'' refers to the name of the application that appears in the list of installed programs on Windows.
		``Autoinstall'' reflects the ability of the installer to automatically proceed with the installation without user interaction (beyond launching the executable and agreeing to the UAC prompt), i.e., it does not require clicking a button first or giving consent.
		``Open webpage'' indicates whether a Wajam website is opened at the end of the installation (typically to congratulate the user).
		``Stealthy'' indicates whether the installation process is totally transparent to the user. It requires Autoinstall and not opening a webpage by the end of the setup, and also not showing any setup window.
		``Rootkit'' indicates the ability to hide the installed application folder from the user.
		Finally, ``Origin'' indicates the provenance of the sample.
	\end{threeparttable}\hspace{0.1in}
	\label{tab:wajam-summary}

\end{table*}

\begin{table*}[t]
	\centering
	\scriptsize
	\caption{List of 332 domains that appear to belong or have belonged to Wajam}
	\vspace{.1in}
	\begin{threeparttable}
		\begin{tabular}{lllll}
			4hewl9m5xz.xyz                                & searchawesome-apps.com               & henaulttechnology.com                     & technologieadrienprovencher.com      & technologiemonroe.com               \\
			4rfgtyr5erxz.com                              & searchesandfind.com                  & hutchisontechnology.com                   & technologiearmandlamoureux.com       & technologiemontorgueil.com          \\
			94j7afz2nr.xyz                                & searchfeedtech.com                   & jarbontechnology.com                      & technologiebarachois.com             & technologiemontroyal.com            \\
			9rtrigfijgu.com                               & searchforall.net                     & jeanlesagetechnology.com                  & technologiebeaubourg.com             & technologiemontrozier.com           \\
			9ruey8ughjffo.xyz                             & searchforfree.net                    & jolicoeurtechnology.com                   & technologiebeaumont.com              & technologiemounac.com               \\
			im1.xyz                                       & searchnewsroom.com                   & kingswoodtechnology.com                   & technologiebellechasse.com           & technologienouaillac.com            \\
			im2.xyz                                       & searchnotifications.com              & kingwintechnology.com                     & technologiebeloeil.com               & technologienullarbor.com            \\
			ta14th1arkr1.xyz                              & search-ology.com                     & labroyetechnology.com                     & technologiebernard.com               & technologieoutremont.com            \\
			wj1.xyz                                       & searchpage.com                       & langeliertechnology.com                   & technologieberri.com                 & technologiepapineau.com             \\
			wj2.xyz                                       & searchpageresults.com                & laubeyrietechnology.com                   & technologieboisseleau.com            & technologiepayenne.com              \\
			wj3.xyz                                       & searchpage-results.com               & launtontechnology.com                     & technologieboissy.com                & technologiepeaches.com              \\
			wj4.xyz                                       & searchpage-results.net               & laurendeautechnology.com                  & technologiebombarderie.com           & technologiepelletier.com            \\
			wj5.xyz                                       & searchsymphony.com                   & lauriertechnology.com                     & technologiebouloi.com                & technologiepiddington.com           \\\cline{1-1}
			autodownload.net                              & searchtech.net                       & mandartechnology.com                      & technologiebourassa.com              & technologiepillac.com               \\
			autotelechargement.net                        & securesearch.xyz                     & manillertechnology.com                    & technologieboussac.com               & technologieprevert.com              \\
			coolappinstaler.com                           & seekoutresultz.com                   & mansactechnology.com                      & technologiebreck.com                 & technologiequainton.com             \\
			customsearches.net                            & social2search.com                    & mercilletechnology.com                    & technologiecalmont.com               & technologierachel.com               \\
			datawestsoftware.com                          & socialwebsearch.co                   & meridiertechnology.com                    & technologiecarmenbienvenue.com       & technologierambuteau.com            \\
			dateandtimesync.com                           & superdownloads.com                   & mertontechnology.com                      & technologiecartier.com               & technologierivolet.com              \\
			dkbsoftware.com                               & supertelechargements.com             & monestiertechnology.com                   & technologiechabanel.com              & technologieruso.com                 \\
			download-flv.com                              & vpn-free.mobi                        & monroetechnology.com                      & technologiechabot.com                & technologierutherford.com           \\
			download-install.com                          & wajam.com                            & montorgueiltechnology.com                 & technologiechamoille.com             & technologiesagard.com               \\
			downloadmngr.com                              & wajam-download.com                   & montroziertechnology.com                  & technologiechamplain.com             & technologiesaintdenis.com           \\
			downloadtryfree.com                           & youcansearch.net                     & mounactechnology.com                      & technologiecharlevoix.com            & technologiesaintdominique.com       \\\cline{2-2}
			downlowd.com                                  & adrienprovenchertechnology.com       & nouaillactechnology.com                   & technologiechaumont.com              & technologiesaintjoseph.com          \\
			downlowd.org                                  & armandlamoureuxtechnology.com        & nullarbortechnology.com                   & technologiechavanac.com              & technologiesaintlaurent.com         \\
			fastappinstall.com                            & barachoistechnology.com              & papineautechnology.com                    & technologiecherrier.com              & technologiesainturbain.com          \\
			fastfreeinstall.com                           & beaubourgtechnology.com              & payennetechnology.com                     & technologiechesterton.com            & technologiesearchawesome.com        \\
			fastnfreedownload.com                         & bellechassetechnology.com            & peachestechnology.com                     & technologieclairavaux.com            & technologiesentier.com              \\
			fastnfreeinstall.com                          & bernardtechnology.com                & pelletiertechnology.com                   & technologiecoloniale.com             & technologiesherman.com              \\
			file-extract.com                              & berritechnology.com                  & piddingtontechnology.com                  & technologiecormack.com               & technologiesirwilfridlaurier.com    \\
			fileextractor.net                             & boisseleautechnology.com             & pillactechnology.com                      & technologiecremazie.com              & technologiesnowdon.com              \\
			fileopens.com                                 & boissytechnology.com                 & plateau-technologies.com                  & technologiecubley.com                & technologiesommery.com              \\
			findresultz.com                               & bombarderietechnology.com            & preverttechnology.com                     & technologiedollard.com               & technologiestdenis.com              \\
			flvplayer-hd.com                              & bouloitechnology.com                 & quaintontechnology.com                    & technologiedrapeau.com               & technologiestlaurent.com            \\
			freeappdownloader.com                         & bourassatechnology.com               & racheltechnology.com                      & technologieduluth.com                & technologiestuart.com               \\
			freeappinstall.com                            & boussactechnology.com                & rambuteautechnology.com                   & technologieemerson.com               & technologietazo.com                 \\
			freeusip.mobi                                 & brecktechnology.com                  & rivolettechnology.com                     & technologieferronnerie.com           & technologieterusse.com              \\
			imt-dashboard.tech                            & calmonttechnology.com                & sagardtechnology.com                      & technologieflagstick.com             & technologiethorel.com               \\
			insta-download.com                            & carmenbienvenuetechnology.com        & saintdominiquetechnology.com              & technologiefullum.com                & technologietofino.com               \\
			install-apps.com                              & cartiertechnology.com                & saintjosephtechnology.com                 & technologiefulmar.com                & technologietoleto.com               \\
			installappsfree.com                           & chabaneltechnology.com               & sainturbaintechnology.com                 & technologiefumier.com                & technologietourville.com            \\
			installateurdappscool.com                     & chabottechnology.com                 & search-technology.net                     & technologiegarfield.com              & technologietravassac.com            \\
			installationdappgratuite.com                  & chamoilletechnology.com              & sentiertechnology.com                     & technologiegarnier.com               & technologietreeland.com             \\
			installationrapideetgratuite.com              & champlaintechnology.com              & shermantechnology.com                     & technologieglencoe.com               & technologietrudeau.com              \\
			installationrapidegratuite.com                & charlevoixtechnology.com             & sirwilfridlauriertechnology.com           & technologiegoyer.com                 & technologieturenne.com              \\
			installeriffic.com                            & chaumonttechnology.com               & snowdontechnology.com                     & technologiegrendon.com               & technologievanhorne.com             \\
			installerus.com                               & chavanactechnology.com               & sommerytechnology.com                     & technologiehenault.com               & technologievanoise.com              \\
			installsofttech.com                           & cherriertechnology.com               & tazotechnology.com                        & technologiehutchison.com             & technologievassy.com                \\
			ios-vpn.com                                   & chestertontechnology.com             & terussetechnology.com                     & technologiejarbon.com                & technologieviau.com                 \\
			main-social2search.netdna-ssl.com             & clairavauxtechnology.com             & thoreltechnology.com                      & technologiejeanlesage.com            & technologievimy.com                 \\
			media-c9hg3zwqygdshhtrps.stackpathdns.com     & colonialetechnology.com              & tofinotechnology.com                      & technologiejolicoeur.com             & technologievouillon.com             \\
			mileendsoft.com                               & cormacktechnology.com                & toletotechnology.com                      & technologiekingswood.com             & technologiewendlebury.com           \\
			notification-results.com                      & cremazietechnology.com               & tourvilletechnology.com                   & technologiekingwin.com               & technologiewilson.com               \\
			notifications-page.com                        & cubleytechnology.com                 & travassactechnology.com                   & technologielabroye.com               & technologiewiseman.com              \\
			notifications-service.info                    & despinstechnology.com                & trudeautechnology.com                     & technologielangelier.com             & technologiewoodham.com              \\
			notifications-service.io                      & drapeautechnology.com                & turennetechnology.com                     & technologielaubeyrie.com             & technologiewoodstream.com           \\
			pagerecherche.com                             & emersontechnology.com                & vanhornetechnology.com                    & technologielaunton.com               & technologiewotton.com               \\
			premiumsearchhub.com                          & ferronnerietechnology.com            & vanoisetechnology.com                     & technologielaurendeau.com            & technologieyvonlheureux.com         \\ \cline{5-5}
			premiumsearchresults.com                      & fullumtechnology.com                 & vassytechnology.com                       & technologielaurier.com               & technologyflagstick.com             \\
			premiumsearchtech.com                         & fulmartechnology.com                 & viautechnology.com                        & technologiemandar.com                & technologyrutherford.com            \\
			result-spark.com                              & fumiertechnology.com                 & videos-conversion.com                     & technologiemaniller.com              & technologytreeland.com              \\
			resultsstream.com                             & garfieldtechnology.com               & vouillontechnology.com                    & technologiemansac.com                & technologywilson.com                \\
			searchawesome.net                             & garniertechnology.com                & wendleburytechnology.com                  & technologiemercille.com              & technologywoodstream.com            \\
			search-awesome.net                            & get-notifications.com                & woodhamtechnology.com                     & technologiemeridier.com              &                                     \\
			searchawesome2.com                            & glencoetechnology.com                & wottontechnology.com                      & technologiemerton.com                &                                     \\
			searchawesome3.com                            & grendontechnology.com                & yvonlheureuxtechnology.com                & technologiemonestier.com             &                                     \\\cline{3-3}
			
		\end{tabular}\par
	\end{threeparttable}
	\label{tab:wajam-domains}
\end{table*}

\section{Anti-analysis and evasion details}
\label{sec:antianalysis-details}

\subhead{Decrypting payloads}
Steganography-based samples D14--18 protect the BRH, by XORing it with a random string found in a stub DLL. Due to the challenges in understanding the decryption routine to find the key, we found that it is easier to brute-force the decryption with all printable strings from that stub DLL until an executable format is decrypted. Alternatively, since parts of the PE headers are predictable, it is possible to recover this key using a known-plaintext attack. However, since D17, this attack is no longer possible as the plaintext is further compressed using a custom method for which there is no known fixed values.

Similarly, the goblin DLL is compressed and encrypted starting from C6 using RC4 and a hardcoded 16-byte key. The key is located in the main executable and can be found by extracting all strings and trying them to decrypt the DLL until a valid GZip header appears. 

Finally, a separate updater runs a Windows service that relies on an encrypted payload called \texttt{service.dat}. In D11--15, the encryption also simply relies on a 16-byte XORed pattern; however, it is not found as plaintext in the main or updater file. Instead, by XORing a known pattern from the PE header, we can recover the key. To fix this weakness, samples starting from D16 switched to RC4, forcing the search of the key obfuscated in one of the executables.

\begin{figure}[h]
	\centering
	\begin{tabular}{lllll}
		\raisebox{-.5\height}{\includegraphics[width=.1\linewidth]{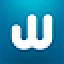}}
		&
		\raisebox{-.4em}{$\cap$}
		&
		\raisebox{-.5\height}{\includegraphics[width=.1\linewidth]{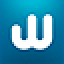}}
		&
		\raisebox{-.4em}{$=$}
		&
		\raisebox{-.5\height}{\includegraphics[width=.1\linewidth]{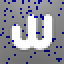}}
	\end{tabular}
	\vspace{-.1in}
	\caption{Icon polymorphism with slight pixel alteration}
	\label{fig:icons-poly}
	\vspace{-.15in}
\end{figure}

\begin{figure}[ht]
	\centering
	\includegraphics[width=.1\linewidth]{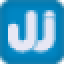}
	\,
	\includegraphics[width=.1\linewidth]{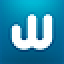}
	\,
	\includegraphics[width=.1\linewidth]{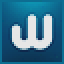}
	\,
	\includegraphics[width=.1\linewidth]{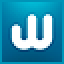}
	\,
	\includegraphics[width=.1\linewidth]{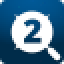}
	\,
	\includegraphics[width=.1\linewidth]{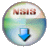}
	\,
	\includegraphics[width=.1\linewidth]{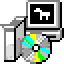}
	\\[3px]
	\includegraphics[width=.1\linewidth]{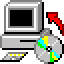}
	\,
	\includegraphics[width=.1\linewidth]{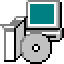}
	\,
	\includegraphics[width=.1\linewidth]{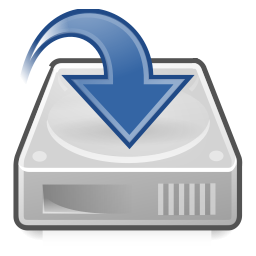}
	\,
	\includegraphics[width=.1\linewidth]{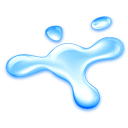}
	\,
	\includegraphics[width=.1\linewidth]{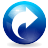}
	\,
	\includegraphics[width=.1\linewidth]{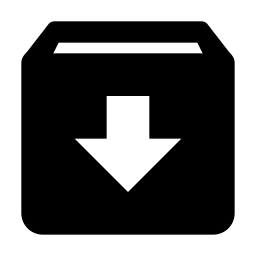}
	\,
	\includegraphics[width=.1\linewidth]{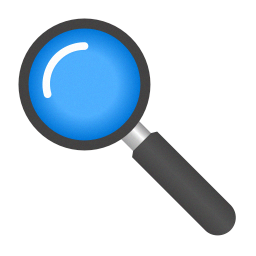}
	\\[3px]
	\includegraphics[width=.1\linewidth]{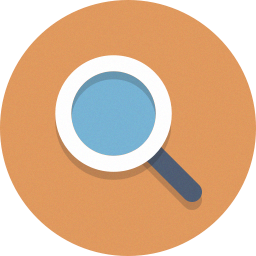}
	\,
	\includegraphics[width=.1\linewidth]{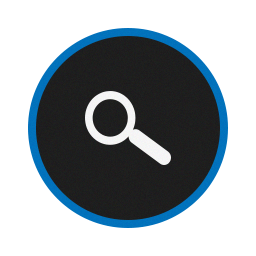}
	\,
	\includegraphics[width=.1\linewidth]{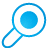}
	\,
	\includegraphics[width=.1\linewidth]{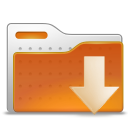}
	\,
	\includegraphics[width=.1\linewidth]{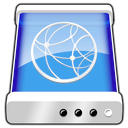}
	\,
	\includegraphics[width=.1\linewidth]{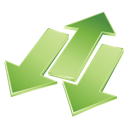}
	\caption{Icons used in the Wajam's installers we collected}
	\label{fig:icons}
	\vspace{-.1in}
\end{figure}

\subhead{T\tIcons: Changing static resources}
Early versions of Wajam shared the same icon on their installers. 
The icon is later changed between variants at few random pixel locations. The color of these pixels is slightly altered to give a new icon while remaining visibly identical, see Figure~\ref{fig:icons-poly}. As a result, the hash of the resource section varies, preventing easy resource fingerprinting.
Starting from D11, Wajam pick random icons from third party icon libraries for both the installer and installed binaries. An illustration is given in Figure~\ref{fig:icons}.

\subhead{T\tNestExec: Nested executables}
From C8, Wajam's main installer unpacks and runs a second NSIS-based installer.

\subhead{T\tPayloadEnc: Payload compression and encryption}
The nested installer is encrypted starting from C10, with the key appended at the end of the ciphertext.
Similarly, the goblin DLL is compressed and encrypted starting from C6 using RC4 and a hardcoded 16-byte key. From D11, the updater is also encrypted with a hardcoded XOR key, then with RC4 in D16.
The injection rules and updates fetched by Wajam are also encrypted (see Section~\ref{sec:flaws}).

\subhead{T\tObfsKey: Obfuscated key reconstruction}
In D17--19, up to two keys are combined and reconstructed from arbitrary string manipulations over the key found in the ciphertext.

\subhead{T\tObfsNSIS: Obfuscated installer script}
The NSIS scripts, which can be decompiled from installers, are obfuscated with thousands of variables and string manipulation operations. We could not find a description of such behavior in the literature. Note that techniques to prevent the identification and recovery of NSIS installers are not used~\cite{nsis-decompile}.
Unlike the nested installer, the outer one remains unobfuscated. This could be done to avoid simple heuristics.

\subhead{T\tObfsNetPS: .NET and Powershell obfuscation}
In the FiddlerCore generation, the Windows service is responsible for adjusting the browser proxy settings and launching the FiddlerCore-based network proxy written in C\#.
Samples from 2014 are not obfuscated and the C\#/.NET components are decompilable.
Starting from sample B4, the method and variable names of C\# components are randomized.
The deobfuscator de4dot~\cite{de4dot} detects that Dotfuscator~\cite{dotfuscator} was used to obfuscate the program; however, only generic method and variable names were reconstructed. Also, de4dot does not remove obvious dead code. Indeed, useful lines of code are interleaved with string declarations made of concatenated random strings. Since such strings are never used, except possibly in the declaration of other such strings, they are easy to remove automatically.

The Powershell persistence module consists of a long \emph{encrypted standard string}, using a user-specific key. As the script runs with SYSTEM privileges, only this account can successfully decrypt the string, revealing another Powershell script that is then invoked. Since decrypting such strings is not directly allowed, the script converts the standard string to a \emph{SecureString}, creates a \texttt{PSCredential} object, and sets the SecureString as the password. Then, it obtains the plaintext password from this object.

\subhead{T\tWhitelist: Auto-whitelisting}
From D5, the installer whitelists the installed program paths in Windows Defender.
Wajam inserts the paths of its main components under {\texttt{HKLM\textbackslash{}Software\textbackslash{}Microsoft\textbackslash{} Windows Defender\textbackslash{}Exclusions\textbackslash{}Paths}}.

\subhead{T\tMRT: Disabling MRT}
From D12, the installer also disables the monthly scans by Windows Malicious Software Removal Tool (MRT) along with the reporting of any detected infections.

\subhead{T\tObfsString: String obfuscation and encryption}
Since C1, string literals in the installed binaries are all XORed with a per-string key.

\subhead{T\tDynamicCalls: Dynamic API calls}
External library calls are made dynamically by calling the \texttt{LoadLibrary} API function provided with a DLL name as argument (obfuscated with T\tObfsString).

\subhead{T\tEncCode: Encrypted code}
The main executable's code section is encrypted in D5--10 with a custom algorithm based on several byte-wise XOR and subtraction operations. Chunks of 456KiB are decoded with the same logic, while each chunk is decoded differently.
Such samples correlate with installers where the file name is prefixed with ``WBE\_crypted\_bundle\_'', suggesting that the encryption layer was added after compilation, possibly by a third-party toolkit.

\subhead{T\tRandomNames: Randomized names}
From B4, installed executable filenames appear random. The installation folder itself becomes randomized from C14 and D3. The names are actually derived from the original name (e.g., wajam.exe), combined with the Machine GUID obtained from registry, and hashed, i.e., \texttt{md5(GUID+filename)}.\footnote{For instance, \texttt{C:\textbackslash{}Program Files\textbackslash{}WaNetworkEn\textbackslash{}wajam.exe} becomes \texttt{C:\textbackslash{}Program Files\textbackslash{}\b686d944556d5de03afc6aa639bff9c7\textbackslash{} 06ca8c13762fca02c5dae8e502fd91c9.exe}, with the folder name corresponding to \texttt{md5(MachineGUID+`WaNetworkEn')} and the filename taken from \texttt{md5(MachineGUID+`wajam.exe')}.}
This pattern is also used in the common name of root certificates from  the fourth generation (see Appendix~\ref{sec:cn-generation}).

\subhead{T\tRootkit: Rootkit}
C11,17,18 rely on a kernel-mode driver to hide the installation folder from the user space, effectively turning Wajam into a rootkit.
C11 also remains even more stealthy as it does not register itself as an installed program and hence does not appear in the list for users to uninstall it.
The file system driver responsible for hiding Wajam's files is called Lacuna and is either named \texttt{pcwtata.sys} or similar, and is signed by DigiCert. 

\subhead{T\tPersistence: Persistence module}
Wajam establishes persistence through executables or scripts that are left in the \texttt{C:\textbackslash{}Windows} folder and not removed by uninstalling the product. While executables could be detected by antiviruses, Wajam leverages (obfuscated) Powershell scripts in samples C17, D3 and D12--13. A scheduled task is left on the system to trigger the persistence module at user logon.
From D14 onward, the persistence module is a regular executable, inheriting some anti-analysis techniques previously mentioned, and set up as a Windows service that starts at boot-time.
The module checks for the presence of the installation directory and main executable. If they do not exist, the module follows the process of updating the application by querying a hardcoded URL to download a fresh variant. This behavior is mostly intended for reinstalling the application after it has been uninstalled, or removed by an antivirus. However, we found that the hardcoded URL is not updated throughout the lifetime of the module on the system, and could be inaccessible when necessary. \looseness=-1

\subhead{T\tDetection: Detection of installed antiviruses}
In every sample since C6, Wajam looks for the presence of a series of 22 major antiviruses and other endpoint security software, then attaches the list of detected products to almost every query it makes to Wajam's server. This might be used to evaluate the distribution of AVs among victims and tailor efforts to evade the most popular ones.
Notably, some of the listed products are intended for business use only, e.g.,
AhnLab and McAfee Endpoint, raising concerns that Wajam might also targets enterprises specifically.

\section{Unique IDs}
Two unique identifiers are generated during installation, and written in the Windows registry. All requests made to Wajam's servers include these identifiers.
The first one, called \texttt{unique\_id} or \texttt{uid} is generated as the uppercased MD5 hash of the combination of: 1) the MAC address of the main network adapter, 2) the path for the temporary folder for applications (which contains the user account's name), and 3) the corresponding disk's serial number. 
The calculation of second identifier, \texttt{machine\_id} or \texttt{mid}, appears to intend including the Machine GUID; however, a programming error fails to achieve this goal, and instead includes some artifact of the string operations performed on the MAC address. In our case, the \texttt{mid} was simply the MAC address prepended by a ``1''. This issue was never fixed.

\section{Updates and injections}
\begin{sloppypar}
Program updates are found in an \texttt{update} or \texttt{manifest} file, generally located at \texttt{/webenhancer/update}, \texttt{/browserenhancer/update} or \texttt{/proxy/manifest} on the remote server.
Similarly, traffic injection rules are called \texttt{injections} or \texttt{mapping} (located at \texttt{/addon/mapping} or \texttt{/webenhancer/injections}).
Finally, the third generation specifically retrieves a \texttt{config} file (\texttt{/webenhancer/config}).
\end{sloppypar}

\subhead{Bootstrap and cache}
The first update is fetched from a hardcoded URL. Later updates are made based on the ``update\_url'' parameter found in the previously fetched file. Once the injection rules are downloaded, they are stored in the program's folder in plaintext in a file named \texttt{WJManifest} for early samples (i.e., B2 and earlier), or encrypted as is in a file named \texttt{waaaghs} or its obfuscated name.
Browser hooking rules are cached similarly, under a file named \texttt{snotlings} or its obfuscated version.

\subhead{Injection methods}
The third generation of Wajam injects a DLL into browser processes, which further hooks a number of functions to manipulate the traffic. While the offsets of the functions are available in the hourly update for Chromium-based browsers, IE and Firefox do not require additional information since the functions to be hooked are readily exported by \texttt{wininet.dll} (in the case of IE) and \texttt{nss3.dll} (for Firefox), and hence can be found easily at runtime. Given the names corresponding to the addresses found in this update file, e.g., \texttt{PR\_Write}, \texttt{SSL\_read\_impl}, Wajam seems to follow the same function hooking strategy to inject content in the network traffic as the Citadel malware~\cite{citadel-virusbulletin}.

Wajam avoids intercepting non-browser applications as evident from a blacklistlist of process names in the update file, e.g., dropbox.exe, skype.exe, bittorrent.exe. Additionally, a whitelist is also present, including the name of supported browser processes; however, it appears not to be used.

Furthermore, Wajam seems to have had difficulties handling certain protocols and compression algorithms in the past. It disables SPDY in Firefox.
Before Chrome version 46, Wajam also modifies the value located at a given offset that represents whether SPDY is enabled to disable this feature. Similarly, the SDCH compression algorithm is disabled.
The number of functions to be hooked evolves from one version of the browser to another, with a different set for 32 and 64-bit versions, sometimes including only \texttt{PR\_}(\texttt{Read}, \texttt{Write}, \texttt{Write\_App}, \texttt{SetError}, \texttt{Close}), or also \texttt{SSL\_read\_impl}.

\begin{figure}[tp]
  \centering
  \footnotesize
  \begin{verbatim}
  [hooks]
    [chrome]
      [...]
      [66_0_3353_2]
        [32bits]
          [PR_Close] => 0x0181C296
          [PR_Write_App] => 0x01824532
          [SSL_read_impl] => 0x01817684
        [64bits]
          [PR_Close] => 0x02318A7C
          [PR_Read] => 0x02312A0C
          [PR_Write] => 0x0232307C
          [PR_Write_App] => 0x0232307C
          [SSL_read_impl] => 0x02312A0C
  \end{verbatim}
  \vspace{-0.15in}
  \caption{Browser injection rule for Chrome 66.0.3353.2}
  \label{fig:inj-browser}
  
\end{figure}

\begin{table*}[tbp]
	\centering
	\scriptsize
	\setlength{\tabcolsep}{3pt}
	\caption{Fingerprints for Wajam-issued leaf certificates (SQL regular expression syntax)}
	\begin{threeparttable}
		\begin{tabular}
			{lp{.5in}ll}
			\textbf{\#} & \textbf{Matches\newline samples} & \textbf{Operator} & \textbf{Issuer Distinguished Name} \\ \hline
			1 & B1--B3   & = & emailAddress=info@wajam.com, OU=Created by http://www.wajam.com, O=WajamInternetEnhancer, CN=Wajam\_root\_cer \\
			2 & B4--B5   & = & emailAddress=info@technologiesainturbain.com, OU=Created by http://www.technologiesainturbain.com, O=WajamInternetEnhancer, CN=WNetEnhancer\_root\_cer \\
			3 & B6       & = & emailAddress=info@technologievanhorne.com, OU=Created by http://www.technologievanhorne.com, O=WajamInternetEnhancer, CN=WaNetworkEnhancer\_root\_cer \\
			4 & D1--D10  & REGEXP & \string^emailAddress=info@technologie.+\textbackslash.com, C=EN, CN=[0-9a-f]\{16\}\$ \\
			5 & D11--D21 & REGEXP & \string^C=EN, CN=[0-9a-f]\{16\} 2\$ \\
			6 & From D22 & REGEXP & \string^C=EN, CN=([YZMNO][WTmj2zGD][FEJINMRQVUZYBAdchglk][h-mw-z0-5])\{2,4\} 2\$ \\
			7 & More recent & REGEXP & \string^C=EN, CN=([YZMNO][WTmj2zGD][FEJINMRQVUZYBAdchglk][h-mw-z0-5])\{1,3\}[YZMNO][WTmj2zGD][FEJINMRQVUZYBAdchglk] 2\$ \\
			8 & More recent & REGEXP & \string^C=EN, CN=([YZMNO][WTmj2zGD][FEJINMRQVUZYBAdchglk][h-mw-z0-5])\{1,3\}[YZMNO][WTmj2zGD] 2\$ \\
			9 & More recent & REGEXP & \string^C=EN, CN=([YZMNO][WTmj2zGD][FEJINMRQVUZYBAdchglk][h-mw-z0-5])\{1,3\}[YZMNO] 2\$ \\
		\end{tabular}\par
	\end{threeparttable}
	\label{tab:wajam-tls-fingerprints}
	
\end{table*}

\section{Root certificate fingerprints}
\label{sec:cn-generation}
\subhead{Common Name generation}
Recovering this algorithm is not straightforward as several intermediate functions separate the CN generation from the certificate generation.
We first identify the function in charge of retrieving the Machine GUID from the registry, and label the parent responsible for concatenating a given string to it and applying the MD5 hash. Then, we identify the function that writes the certificate to a file named after the CN, and trace the origin of the filename to a function that calls the previously labeled function. The argument passed in the call corresponds to the concatenated string.
After observing in a few samples that the concatenated string matches the registry key of the installed application, we simply proceed to try this key to match the generated certificates in other samples. The various application names can be found in Table~\ref{tab:root-certs}.

In the last two samples (D22--23), the process is similar; however, only the 12 first hexadecimal characters of the MD5 hash are taken into account, which are further encoded using base64 giving e.g., \texttt{ZmJiYmRiODYxNTZi}.
We also found that samples branded as SearchAwesome install a certificate with a CN appended with the digit ``2'', corresponding to a new feature in ProtocolFilters that appeared in May 2015~\cite{netfilter-changelog}.

\subhead{Fingerprints}
Table~\ref{tab:wajam-tls-fingerprints} shows the regular expressions to match Wajam's 2nd and 4th generation root certificate Distinguished Names (DN) based on our observations.

While the first 3 DNs are static, others capture all possible combinations which we reverse-engineered from Wajam's binaries. In particular, patterns 4--5 match a CN that represents 16 hexadecimal characters, thus this type of CN caries $\log_2(16^{16})=64$ bits of entropy.
Patterns 6--9 correspond to samples where the hexadecimal CN is base64-encoded and truncated at various lengths. Due to the limited space of hexadecimal characters to encode, the resulting CN follows a repeated pattern of 4 letters from different sets, e.g., the first encoded letter can only be an Y, Z, M, N or O. Not all combinations of letters from the sets are possible, thus these patterns are overestimating possible fingerprints. Pattern 6 can match up to 16 characters, which translates into 12 hexadecimal characters and thus 48 bits of entropy.

During our scans in Mar.\ 2019, we also found certificates with similar fingerprints as produced by D22 and D23; however, their issuer CN were shorter. When we detected such cases, we also fetched the web page and found that the injected content also points to Wajam domains. Since samples from Mar.\ 2019 we could obtain from the known distribution URL do not generate such certificates, it could be possible that we are missing another ``branch'' of Wajam.
For instance, the shortened CN ``MDM5Z 2'' caries 12 bits for the first four letters + $2.32$ bits for the 5th character (one out of five), resulting in an overall entropy of $14.32$ bits.

\end{document}